\documentclass[12pt]{article}

\usepackage{graphicx}
\usepackage{epstopdf}
\usepackage[body={17.5cm, 21cm},right=2cm]{geometry}
\usepackage{amssymb}
\usepackage{amsmath}

\newcommand\eea{\end{eqnarray}}
\newcommand\bea{\begin{eqnarray}}
\newcommand{\sfrac}[2]{{\textstyle\frac{#1}{#2}}}
\newcommand\di{\partial}

\def\la{\langle}
\def\ra{\rangle}
\def\beq{\begin{equation}}
\def\eeq{\end{equation}}
\def\d{\partial}

\def\d{\partial}
\def\l{\left(}
\def\r{\right)}
\def\la{\langle }
\def\ra{\rangle }
\newcommand{\be}{\begin{equation}}
\newcommand{\ee}{\end{equation}}
\newcommand{\ba}{\begin{align}}
\newcommand{\ea}{\end{align}}
\newcommand{\bg}{\begin{gather}}
\newcommand{\eg}{\end{gather}}
\newcommand{\bseq}{\begin{subequations}}
\newcommand{\eseq}{\end{subequations}}
\newcommand{\tg}{\mathop{\rm tg}\nolimits}

\begin{document}
\setcounter{page}{0}
\thispagestyle{empty}

\begin{titlepage}

\begin{flushright} {\footnotesize IC/2008/002}  \end{flushright}

\begin{center}

\vspace{1.cm}

{\LARGE \sc{
The Phase Transition\\ 
\vspace{.1cm}
to \\
\vspace{.5cm}
Eternal Inflation}}\\[1.5cm]
{\large Paolo Creminelli$^{\rm a}$, Sergei Dubovsky$^{\rm b,c}$, Alberto Nicolis$^{\rm d}$, \\[.2cm] Leonardo Senatore$^{\rm b}$, and Matias Zaldarriaga$^{\rm b,e}$}
\\[.8cm]

{\small \textit{$^{\rm a}$
Abdus Salam International Centre for Theoretical Physics, \\ 
Strada Costiera 11, 34014 Trieste, Italy}}

\vspace{.2cm}
{\small \textit{$^{\rm b}$ 
Jefferson Physical Laboratory,  Harvard University, \\ 
Cambridge, MA 02138, USA}}

\vspace{.2cm}
{\small \textit{$^{\rm c}$ 
Institute for Nuclear Research of the Russian Academy of Sciences, \\
60th October Anniversary Prospect, 7a, 117312 Moscow, Russia}}
        
\vspace{.2cm}
{\small \textit{$^{\rm d}$ 
Department of Physics and ISCAP,\\
 Columbia University, 
New York, NY 10027, USA }}

\vspace{.2cm}
{\small \textit{$^{\rm e}$  
Center for Astrophysics, Harvard University, \\
Cambridge, MA 02138, USA}}

\end{center}

\vspace{.8cm}
\begin{abstract}

For slow-roll inflation we study the phase transition to the eternal regime. Starting from a finite inflationary
volume,
we consider the volume of the universe at reheating as order parameter. We show that there exists a critical value
for the classical inflaton speed, $\dot\phi^2/H^4 = 3/(2 \pi^2)$, where the probability distribution for the reheating volume undergoes a sharp
transition. In particular, for sub-critical inflaton speeds all distribution moments become infinite. We show
that at the same transition point the system develops a non-vanishing probability of having a strictly infinite
reheating volume, while retaining a finite probability for finite values.
Our analysis represents the exact quantum treatment of the system at
lowest order in the slow-roll parameters and $H^2/M_{\rm Pl}^2$.

\end{abstract}

\end{titlepage}

%%%%%%%%%%%%%%%%%%%%%%%%%%%%%%%%%

% \section{Introduction}

%%%%%%%%%%%%%%%%%%%%%%%%%%
%%				ALBERTO			%%
%%%%%%%%%%%%%%%%%%%%%%%%%%

\section{Introduction and setup\label{sec:intro}}

\subsection{Heuristic thoughts}
Consider a standard slow-roll inflationary model. At the classical level, the inflaton field $\phi$ gently rolls down its potential, and as long as the slow-roll conditions are met the universe inflates. Eventually, once the inflaton kinetic energy becomes comparable to the potential energy, inflation ends and the universe reheats. At the quantum level, on top of this classical story there are small scalar quantum fluctuations, which after re-entering the horizon during the post-inflationary era will lead to the familiar density perturbations.
In the gauge in which the inflaton follows its unperturbed classical trajectory, $\delta \phi=0$, the scalar fluctuations are conveniently parameterized by the variable $\zeta$, which measures the spatial curvature of constant-$\phi$ hypersurfaces, $^{(3)}{\cal R}= -4 \nabla^2 \zeta/a^2$. For a given inflationary history $\big( \phi(t),H(t) \big)$ the quadratic action for scalar perturbations is simply (see e.g.~ref.~\cite{maldacena})
\be
S_{\zeta} = \int \! dt d^3 x \, \sqrt{-g} \:  \frac{\dot \phi^2}{H^2} \, \sfrac12 \big( \di \zeta \big) ^2  \; ,
\ee
where the index contraction is done with the background FRW metric. 

Upon canonical quantization of the above action, the typical size of quantum fluctuations at horizon scales is
\be \label{zeta}
\langle \zeta^2 \rangle_H \sim  \frac{H^4}{\dot \phi^2} \sim \frac1 \epsilon \, \frac {H^2}{M_{Pl}^2} \; ,
\ee
where $\epsilon$ is the usual slow-roll parameter, $\epsilon \equiv \dot \phi^2 / (2 H^2 M_{Pl}^2)$.
$\zeta$ then remains constant outside the horizon, and at horizon re-entering the scalar perturbations show up as density perturbations, with initial amplitude given by $\zeta$ itself,  $\delta \rho / \rho \sim \zeta$. 

From a geometric viewpoint, the variable $\zeta$ is directly the dimensionless perturbation in the spatial metric of constant-$\phi$ hypersurfaces, $g_{ij} = a^2(t) (1+2 \zeta) \, \delta_{ij}$. Hence, small $\zeta$ fluctuations means that these hypersurfaces are slightly curved, whereas the extreme case of $\zeta$ of order one corresponds to a highly deformed configuration. 
From eq.~(\ref{zeta}) we see that the latter case arises when the inflaton potential is so flat as to make $\dot \phi^2$ of order $H^4$, or smaller. In terms of the slow-roll parameter $\epsilon$, this happens when $\epsilon \lesssim H^2/{M_{Pl}^2}$.
This regime corresponds to eternal inflation. Indeed, to see this it is convenient to switch to a different gauge, for example a spatially flat gauge, where scalar perturbations are parametrized through the inflaton fluctuation $\delta \phi$. Then, at lowest order in slow-roll parameters, the quadratic action is that of a minimally coupled, canonically normalized scalar in a fixed background geometry (see for example \cite{maldacena}),
\be \label{deltaphi}
S_{\delta \phi} = \int \! dt d^3 x \, \sqrt{-g} \: \sfrac12 \big( \di \delta \phi \big) ^2  \; ,
\ee
whose typical quantum fluctuations at scales of order $H$ are
\be \label{quantum}
\langle (\delta \phi)^2 \rangle_H \sim H^2 \; .
\ee
On the other hand, the inflaton vev in one Hubble time advances by an amount
\be \label{classical}
\Delta \phi_{\rm cl} = \dot \phi \cdot H^{-1} \; .
\ee
In the ordinary, non-eternal regime quantum fluctuations are just a small correction to the classical story. But if the classical $\dot \phi$ becomes very small---smaller than $H^2$---the classical inflaton advancement (\ref{classical}) drops below the typical size of quantum fluctuations. At that point the classical downhill drift becomes unimportant and the system is dominated by quantum diffusion. In particular backward diffusion is as likely as the forward one. Any given observer will eventually experience reheating---because in a diffusion process the probability for a given particle to touch any given position is always one. Nevertheless inflation is globally eternal, in that backward diffusion, through the exponential expansion of the universe, leads to the creation of more and more Hubble patches which then have a chance themselves to diffuse, and there will always be some Hubble patch that keeps inflating.
Notice that, as mentioned above, the eternal inflation regime sets in precisely when $\zeta$ fluctuations become of order one, signaling large deformations in the geometry. This condition was derived first in  \cite{Linde:1986fd,Goncharov:1987ir}.

The purpose of the present paper is to make progress in the {\em quantitative}  understanding of eternal inflation. Much work has been
done in this direction since the first discovery of this phenomenon \cite{Guth:1980zm,Linde:1982ur,Steinhardt:1982kg,Vilenkin:1983xq,Linde:1986fd} long time ago, and some of the material, especially in this Section, will not be completely
original (see two recent summaries by Guth \cite{Guth:2007ng} and by Linde \cite{Linde:2004kg}  and references therein). The main point will be to find a precise definition of slow-roll eternal inflation (if it exists) and in particular the precise inequality that must be satisfied by the parameters of the model to have eternal inflation.

Such a plan seems to be doomed to failure from the start. For one thing, when the system is in or close to the eternal inflation regime, quantum fluctuations are so large as to overcome the classical dynamics, thus impairing {\em (a)} the semiclassical approximation we started with, and {\em (b)} the approximate homogeneity and isotropy of the universe. Once these two features are lost, there seems to be very little hope to make any quantitative progress.
Indeed, the homogeneity issue is what usually makes so confusing the global picture of false-vacuum eternal inflation, where bubbles of the new vacuum with different cosmological constant keeps forming, giving rise to an intractable spacetime.

However, let's look at these problems more closely for our slow-roll case. As to point {\em (a)} above, large quantum fluctuations do not necessarily mean that the system is intractable. Hardly anybody would suggest that a free scalar in Minkowski space in its vacuum state is intractable, even though its dynamics is all in quantum fluctuations. Large quantum fluctuations are only a problem if they are associated with strong coupling, that is if they correspond to some interactions becoming large, thus leading to the breakdown of perturbation theory. In other words, the issue of large fluctuations is a dynamical question rather than a kinematical one. As long as interactions are small, a quantum field can be reliably dealt with through the perturbative expansion. 
As to point {\em (b)}, although in the eternal inflation regime the typical size of $\zeta$ fluctuations at horizon scales is of order one, we must recall that $\zeta$ parameterizes the scalar deformations in the spatial geometry induced on the equal-$\phi$ surfaces. A large $\zeta$ means that these surfaces have a large intrinsic curvature. This however can arise even if the four-dimensional geometry is virtually unperturbed, but the $\phi$ fluctuations are so large as to dramatically bend the equal-$\phi$ surfaces in the surrounding FRW geometry. In this case the full 4D geometry would still be approximately homogeneous and isotropic, thus allowing once again for a perturbative approach.

% in two qualitatively distinct cases (and in all intermediate combinations): either the full four-dimensional geometry is highly perturbed with respect to the background FRW solution, and this deformation gets projected onto the equal-$\phi$ surfaces, or the four-dimensional geometry is virtually unperturbed, but the $\phi$ fluctuations are so large as to dramatically bend the equal-$\phi$ surfaces in the surrounding FRW geometry. The latter possibility is obviously harmless: in that case the full 4D geometry would still be approximately homogeneous and isotropic, thus allowing once again for a perturbative approach. 

Fortunately {\em in slow roll eternal inflation both perturbative expansions are allowed}, and are in fact one and the same, as we now explain.

\subsection{Why the system is perturbative\label{sec:perturbative}}
To see that the geometry is indeed very close to that of the unperturbed inflationary solution, it is convenient to work in a gauge in which the metric fluctuations are manifestly small.
This is the case for example in the gauge where the scalar perturbation is parametrized by $\delta \phi$ and the equal-time hypersurfaces are flat
\footnote{We are restricting our analysis to  scalar fluctuations. The inclusion of tensor modes is as straightforward as uninteresting: nothing dramatic happens to them in the eternal inflation regime, their amplitude being always of order $H/M_{\rm Pl}$.},
\be
g_{ij} = a^2(t) \delta_{ij} \; .
\ee
Then the other components of the metric at linear order are perturbed by \cite{maldacena}
\bea
&& \delta g_{00} = \frac{\dot \phi}{H M_{\rm Pl}^2} \, \delta \phi \sim   \sqrt{\epsilon} \, \frac{H} {M_{Pl}} \; , \label{metric1} \\ 
&& \delta g_{0i} = - \frac{\dot \phi^2}{2 H^2 M_{\rm Pl}^2} \frac{a^2(t)}{\nabla^2}\nabla_i  \frac{d}{dt} 
\bigg( \frac{H}{\dot \phi} \, \delta \phi \bigg) \sim a(t) \cdot \sqrt{\epsilon} \, \frac{H} {M_{Pl}} \; , \label{metric2}
\eea
where we used the typical size of $\delta \phi$ fluctuations at horizon scales, eq.~(\ref{quantum}).
We explicitly see that by making  $\epsilon$ smaller and smaller, we
are actually {\em suppressing} the geometry fluctuations \cite{Goncharov:1987ir}. In the limit of vanishing $\epsilon$ the spacetime geometry looks unperturbed, but this is an artifact of the linear approximation. At quadratic order there are contributions to $g_{00}$, $g_{0i}$ of order $(H/M_{\rm Pl})^2$, which dominates over the linear ones for $\epsilon \lesssim H^2/M_{\rm Pl}^2$---precisely in the eternal inflation regime. Higher orders are however further suppressed by higher powers of $H^2/M_{\rm Pl}^2$, thus allowing for a perturbative expansion in such a quantity.
As suggested above, the fact that in the eternal inflation regime $\zeta$ blows up is an artifact of the strong bending of the equal-$\phi$ surfaces in an otherwise smooth four-dimensional geometry.

Similarly, for the interactions, in the slow-roll approximation the
leading interactions come from couplings with and within the gravity
sector, while the ones coming from the potential are subdominant \cite{maldacena}.  Then it is clear that as long as metric perturbations are small, interactions will be small as well and the perturbative expansion will be valid.
Indeed, after integrating out the metric fluctuations (\ref{metric1}, \ref{metric2}) in the same gauge as above, one gets trilinear,  two-derivative self-couplings for $\delta \phi$, schematically \cite{maldacena}
\be
S_{\rm 3} \sim \int \frac{\dot \phi}{H M_{\rm Pl}^2} \, \delta\phi \, \di \delta \phi \, \di \delta\phi \; .
\ee
By using again the typical size of inflaton fluctuations, eq.~(\ref{quantum}), we see that at scales of order $H$ interactions are suppressed with respect to the quadratic Lagrangian (\ref{deltaphi}) by
\be
\label{eq:S3}
\frac{S_{\rm 3}}{S_{\delta\phi}} \sim \sqrt{\epsilon} \, \frac{H}{M_{\rm Pl} } \; .
\ee
That  is, for smaller and smaller  $\epsilon$ the system becomes more and more perturbative.
Once again, past the critical value $\epsilon_c \sim H^2/M_{\rm Pl}^2$ new interactions start dominating, namely quartic, two-derivative couplings \cite{Seery:2006vu}
\be
\label{eq:S4}
S_4 \sim \int \frac{1}{M_{\rm Pl}^2}  \, \di^2 \, \delta \phi^4 \sim \frac{H^2}{M_{\rm Pl}^2} \cdot S_{\delta\phi}\; ,
\ee
while higher-order interactions are suppressed by higher powers of $H^2/M_{\rm Pl}^2$.

In summary, in the slow-roll regime geometry fluctuations and interactions are both perturbative,
with expansion parameters $\sqrt{\epsilon} \cdot H/M_{\rm Pl}$ and $H^2 / M_{\rm Pl}^2$ in both cases. These expansion parameters remain small even in the eternal inflation regime $\zeta \sim H/(M_{\rm Pl} \sqrt{\epsilon}) \sim 1$.
Also for $\epsilon \to 0$ the background FRW solution tends to de Sitter space. We are therefore led to the conclusion that in this gauge {\em at lowest order in slow-roll and $H^2 / M_{\rm Pl}^2$ the system is equivalent to a minimally coupled, free scalar in a background de Sitter space, with no dynamical gravity}.
In hindsight, this conclusion is obvious. In the limit in which the potential is flat the background geometry is that of de Sitter space and the only interactions are gravitational in nature, thus leading to effects that are suppressed by $G_N$ times the appropriate power of the only energy scale in the problem, $H$. 
In the following we are going to neglect all corrections in the slow-roll parameters and in $H/M_{\rm Pl}$.
In particular we are going to treat as constant the slowly varying parameters $H$ and $\dot\phi$; towards the end we will come back to discuss what happens when this approximation is relaxed.

Of course the simplification we discussed only applies to the inflationary era. Once inflation ends the slow-roll approximation breaks down, and $\zeta$ fluctuations get converted into density fluctuations. For $\zeta \sim 1$, one gets $\delta\rho/\rho \sim 1$ when the modes come back in the horizon, giving rise to a highly inhomogeneous four-dimensional geometry. Therefore in the eternal inflation regime, $\zeta \gtrsim 1$, the post-inflationary era is as inhomogenous and as confusing as the usual large-scale structure of false-vacuum eternal inflation; but as long as we stick to the inflationary phase our picture of a free scalar in de Sitter space applies. 

%More precisely, $\zeta$ parameterizes the curvature of hypersurfaces of constant energy density $\rho$, which now varies by order one in one Hubble time. Since energy density is what curves the four-dimensional geometry, a large intrinsic curvature in the constant-$\rho$ hypersurfaces now translates into a highly inhomogeneous four-dimensional geometry. Therefore in the eternal inflation regime, $\zeta \gtrsim 1$, the post-inflationary era is as inhomogenous and as confusing as the usual large-scale structure of false-vacuum eternal inflation; but as long as we stick to the inflationary phase our picture of a free scalar in deSitter space applies. 

\begin{figure}[t!]
\begin{center}
\includegraphics[width=14cm]{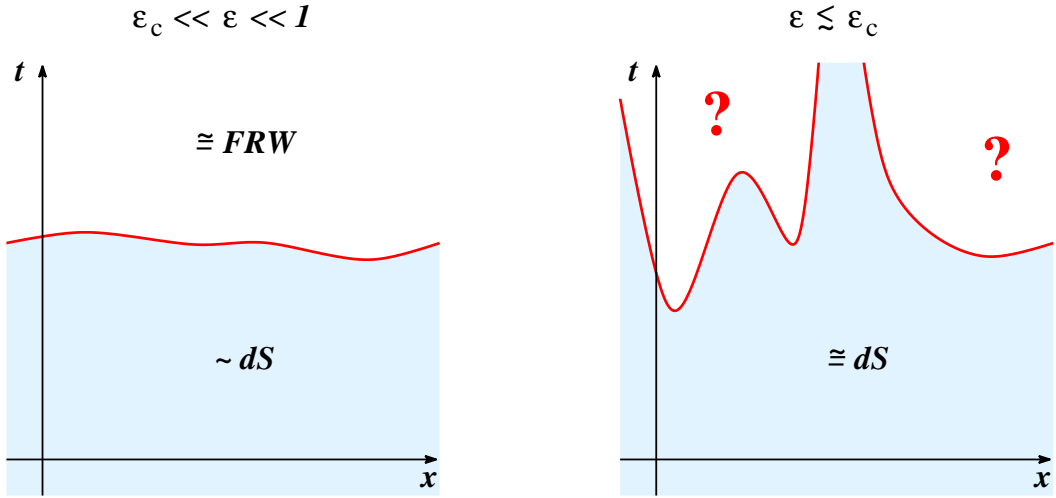}
\caption{\label{reheating} \small \it {\em Left:} For standard, non-eternal inflation the inflationary phase {\em (shaded region)} is nearly de Sitter, with slow-roll corrections and small inhomogenities; the reheating surface {\em (red curve)} is slightly bent; the post inflationary phase is nearly FRW, with small perturbations inherited from the reheating surface. {\em Right:} In the eternal inflation regime, the inflationary phase is very well approximated by de Sitter geometry, with corrections of order $H^2/M_{\rm Pl}^2$, but the reheating surface has wild fluctuations; as a consequence the post-inflationary phase is highly inhomogeneous, and intractable.}
\end{center}
\end{figure}

In any realistic inflationary scenario the breakdown of the slow-roll regime and the consequent reheating of the universe will be gradual processes, but for our purposes it is more convenient to picture them as instantaneously happening at some critical field value, $\phi = \phi_r$. In this approximation reheating corresponds to some definite hypersurface in spacetime.
In the ordinary, non-eternal regime, $\dot \phi/H^2 \gg 1$, the spacetime region before the reheating surface has a nearly de Sitter geometry, with small slow-roll corrections, whereas after the reheating surface the universe is well approximated by a FRW solution with small inhomogenities, whose amplitude is given $\zeta \sim H^2/\dot \phi$. If we now make the ratio $\dot \phi/H^2$ smaller and smaller, the two above approximate descriptions tend to separate: the de Sitter approximation before reheating becomes more and more precise up to the ultimate accuracy of $(H/M_{Pl})^2$, while the FRW one after reheating becomes strongly inadequate past $\dot \phi / H^2 \sim 1$. The situation is schematically depicted in fig.~\ref{reheating}. 
In the eternal inflation regime we have no hope to quantitatively study the large-scale structure of the universe after reheating.
However, we can study the geometry of the reheating hypersurface by approaching it from the de Sitter phase, where our approximations are well under control. We want to sharply characterize eternal inflation through some geometric property of the reheating surface.

\subsection{Finite comoving volume and UV smoothing\label{sec:smoothing}}
A natural candidate for characterizing eternal inflation is the volume of the reheating surface: if starting with a finite inflationary volume at $t=0$, the reheating surface ends up having infinite volume, we are in eternal inflation. As explained above, this should happen below some critical value of $\dot \phi/H^2$, which in our approximations is the only dimensionless parameter of the model. Of course the volume of the reheating surface has statistical fluctuations, which supposedly  are very large close to the eternal inflation regime, so our goal is to determine whether there is a sharp transition value of $\dot \phi/H^2$ at which the statistical properties of the reheating volume abruptly change.

Before studying the volume statistics, we have to define our system more precisely. First, as we already mentioned, it is better to talk about a finite initial inflationary volume, otherwise the reheating volume would be infinite to begin with. Let's therefore consider an inflationary universe compactified on a three-torus, with fixed comoving size $L$. $L$ then plays the role of an infrared cutoff for the inflaton Fourier modes. We will explicitly see that our results do not depend on the value of $L$.
For simplicity we will assume that at the initial time $t=0$, the size of the universe is much larger than the Hubble radius, $L \gg H^{-1}$ (we are choosing $a(t)$ to be unity at $t=0$). This is just for technical convenience, to use integrals rather than sums in Fourier space.

Then, we want to study the constant-$\phi$ hypersurfaces, in particular that with $\phi = \phi_r$. At short distances these surfaces are arbitrarily irregular, because the inflaton fluctuations diverge at shorter and shorter scales, $\langle (\delta \phi)^2 \rangle_k \sim k^2$. This leads to a UV-divergent volume for the equal-$\phi$ surfaces. However this is true even in Minkowski space, and it has nothing to do with the inflationary background we are studying. We can then cutoff the high momenta above some UV scale by smoothing out the field in space over the corresponding UV length scale. Notice that a fixed UV cutoff in physical distances looks time-dependent---actually exponentially shrinking---in comoving coordinates.
We thus consider the smoothed inflaton field
\be
\label{eq:smooth}
\phi_\Lambda (\vec x, t)= \int \! d^3 r \; f_\Lambda (r) \, \phi(\vec x +\vec r / a(t), t ) \; ,
\ee
where $f_\Lambda (r)$ is a smooth function peaked at the origin and with typical width given by $1/\Lambda$, like for instance a Gaussian with $\sigma = 1/\Lambda$, and the explicit factor of $a(t)$ gives the correct time-dependence. Knowing the correlation functions for $\phi$---which in the free field limit we are working in are all determined by just the two-point function---we can compute all the correlation functions for $\phi_\Lambda$. In particular, $\phi_\Lambda$ too will be a Gaussian field, being a linear superposition of Gaussian fields.
It will be technically more convenient for us to work with a sharp cutoff in momentum space and in this case we will see that our results do not depend on the exact size of the cutoff. This strongly suggests that it does not matter how we choose to implement the cutoff, i.e.~what filter function $f_\Lambda (r)$ we pick. For reasons that will soon become clear we need to smooth out the inflaton over several Hubble volumes, that is, we need $\Lambda^{-1}$ to be comfortably larger than $H^{-1}$.

%We will explicitly show that our results will not depend on $\Lambda$, i.e.~on what we choose as the UV cutoff. This also indicates that it does not matter how we choose to implement the cutoff, i.e.~what filter function $f_\Lambda (r)$ we pick---in fact, 
We then concentrate on the constant-$\phi_\Lambda$ surfaces. These are smooth by definition, with no wrinkles below the UV length-scale
$1/\Lambda$. Studying these smoothed-out surfaces rather than the original, `raw' ones also makes more sense as far as reheating is concerned. Indeed, above we were suggesting that reheating happens when the local inflaton vev reaches the critical value $\phi_r$, but of course `local' has to be understood in a smoothed sense---roughly speaking, what matters for reheating should be the inflaton average over at least one Hubble volume.
Once again, the fact that our results will be independent of
$\Lambda$, for $\Lambda \ll H$, gives us confidence that the precise prescription we choose for defining the reheating surface does not matter. 

In the following we will drop the subscript `$\Lambda$' from the smoothed field $\phi_\Lambda$, and the smoothing will be understood unless otherwise stated.

\subsection{Space-likeness of the reheating surface}
Our smoothing procedure is also important for making the reheating surface space-like.
Indeed, when we talk about computing the volume of the reheating
surface, we are implicitly assuming that such a surface is space-like,
so that its volume can be interpreted as the volume of the universe at
the end of inflation. If such a surface were not space-like
everywhere, the `end of inflation' would not be an event in time, but
rather in some space-time region it would look like a boundary in
space. Also, the induced metric on the surface would not be
positive-definite everywhere.  The space-likeness of the reheating
surface is not obvious when quantum fluctuations are taken into
account. In the classical limit the reheating surface is obviously
space-like, as the classical motion gives a $\dot\phi$ with definite
sign. Quantum corrections add an additional motion to the classical
one, as more and more modes enter in the filter $f_\Lambda(r)$ of
eq.~(\ref{eq:smooth}). It is not obvious that the resulting motion
gives rise to a space-like reheating surface \footnote{For a generic inflaton value, different from the reheating one, it is not even clear whether it makes sense to talk about `surface of constant inflaton value'. Taking into account quantum fluctuations in the direction opposite to the classical motion, it is possible to go through the same value of the inflaton many times, so that the points of constant inflaton value will not form a smooth manifold. This cannot happen for the reheating surface, because by definition after crossing $\phi_r$ inflation ends and the system cannot fluctuate back again.}.

Fortunately the smoothing procedure helps in this respect:  the probability for a constant-$\phi$ surface {\em not} to be space-like somewhere is suppressed by $\sim e^{- H^2/\Lambda^2}$, where $\Lambda \ll H$  is our UV cutoff. To see this, consider the inflaton field $\phi$, without any smoothing filter. The equal-$\phi$ surfaces are space-like if and only if $\phi$ itself is a good time variable, that is if and only if $\di_\mu \phi$ is time-like everywhere,
\be \label{timelike}
(\di \phi)^2 > 0
\ee
(we use the $(+,-,-,-)$ metric signature).
Of course $\phi$ is a quantum field, and the above condition cannot hold as an exact operator statement: at any space-time location there is always a finite probability for $\di_\mu \phi$ to point in a space-like direction. The best we can hope for is to make  the above inequality valid as an expectation value, and quantum fluctuations around it small; this issue is general, and it is not peculiar to the eternal regime we want to study.
As to the expectation value, if we split the field into background plus fluctuations, we have
\be \label{expectation}
\langle (\di \phi)^2 \rangle= \dot \phi^2 + \langle (\di \delta \phi)^2 \rangle  \; ,
\ee
where $\dot \phi$ is the classical inflaton speed, and $\delta \phi$ in our approximations is well described by a free, minimally coupled scalar in de Sitter space. The fluctuating piece $\langle (\di \delta \phi)^2 \rangle$ is UV divergent; if we cut it off at some physical momentum $\Lambda$ we get 
\be \label{dphisquare}
\langle (\di \delta \phi)^2 \rangle = - \int^{\Lambda \cdot a(t)}\frac{d^3 k}{(2 \pi)^3} \; \frac1{a^2(t)} \frac{H^2}{2 k} = - \frac{H^2}{8 \pi^2} \Lambda^2 \; ,
\ee
which is negative. We thus see that quantum fluctuations tend to violate eq.~(\ref{timelike}), and that the equal-$\phi$ surfaces cannot be assumed to be space-like at arbitrarily short scales. Only for length-scales larger than the critical UV cutoff
\be \label{critical_cutoff}
\Lambda_c ^{-1} \sim \frac{H}{\dot \phi} = H^{-1} \cdot \frac{H^2}{\dot \phi}
\ee
does the classical piece in eq.~(\ref{expectation}) dominate over the quantum one. We stress again  that this conclusion also applies to standard, non-eternal inflation. For instance, at scales shorter than the above critical cutoff there is no sense in which one can use $\phi$ itself as a clock and define the equal-time surfaces by the gauge-fixing $\delta\phi=0$.  Notice however that in the non-eternal regime $\dot \phi/H^2$ is much larger than one and so the critical cutoff is parametrically smaller than the Hubble radius; on the other hand close to or in the eternal regime $\dot \phi/H^2$ is of order one, thus making the critical cutoff of order of the Hubble radius. This is why in our smoothing procedure we choose to filter out all modes with wavelengths smaller than $\Lambda^{-1} \gg H^{-1}$.

Similarly for the standard deviation of eq.~(\ref{timelike}), at scales larger than $1/\Lambda_c$ the dominant contribution comes from cross-product terms,
\be \label{stdev}
\sqrt{\big \langle \big[ (\di \phi)^2 - \langle(\di \phi)^2 \rangle \big]^2 \big\rangle} 
\sim \sqrt{\dot\phi^2 \, \langle (\delta \dot \phi)^2 \rangle} \sim \dot \phi \cdot \Lambda^2 \; ,
\ee
%\be \label{stdev}
%\sqrt{\big \langle \big[ (\di \phi)^2 - \langle(\di \phi)^2 \rangle \big]^2 \big\rangle} 
%\sim \sqrt{\dot\phi^2 \, \langle (\di \delta \phi)^2 \rangle} \sim \dot \phi \cdot H \Lambda \; ,
%\ee
Requiring this to be negligible compared to the classical value $\dot\phi$ gives a constraint on $\Lambda$: $\Lambda < \Lambda_c \simeq \dot\phi^{1/2}$, which is stronger than (\ref{critical_cutoff}). In the eternal inflation regime however, $\dot\phi \sim H^2$, so that this just reiterates the conclusion 
that the cutoff must be parametrically smaller than $H$. 
\footnote{Eq.~(\ref{stdev}) is the correct expression in the eternal case, where we must choose $\Lambda \ll H$. In the non-eternal case, if we allow for $\Lambda \gg H$ there is another competing term of the form $\Lambda^4$. Anyway one ends up with the constraint $\Lambda < \Lambda_c \simeq \dot\phi^{1/2}$ which is parametrically shorter than the Hubble radius in the non-eternal regime.}

In summary, if we smooth out the inflaton field over length-scales
somewhat larger than the Hubble radius, we can safely assume that the
constant-$\phi$ hypersurfaces are space-like \footnote{\label{freivogel}One may argue
  that, although the smoothed reheating surface is spacelike, the real one may
  be not, as reheating will be sensitive to scales of order $H^{-1}$
  and not much larger. This would be problematic at it implies the
  existence of inflating points in the future light-cone of points
  where inflation already finished. As we discussed, the
  post-inflationary era is not under control and it would therefore be
impossible to proceed. However, a model in which a potential 
giving eternal inflation abruptly ends and reheats is an
idealization. One should consider a realistic potential where $\zeta
\ll 1$ close to reheating. As we will discuss later, this period of
non-eternal inflation does not change our conclusions, but it
introduces a physical smoothing of the reheating surface, as at short
scales $\zeta$ is very small. This implies that the reheating surface
{\em is} space-like. This is the most physical way of looking at our
smoothing procedure: the effect of a final period of inflation during
which quantum fluctuations are small. We thank Ben Freivogel and Raman
Sundrum for discussions about this point.}. Of course there is always a non-vanishing probability of having a large quantum fluctuation that locally brings $\di_\mu \phi$ onto a space-like direction. However thanks to the gaussianity of the inflaton fluctuations, such probability is exponentially small with a width given by eq.~(\ref{stdev}), roughly
\be
P\big( (\di \phi)^2 < 0 \big) \sim e^{-\frac {\dot \phi^2}{\dot \phi \Lambda^2}}  \sim e^{-H^2/\Lambda^2} \; .
\ee
We stress again that the same arguments hold in the non-eternal
regime: the gauge fixing $\delta\phi =0$ only make sense with a
sufficiently small cutoff $\Lambda$ and neglecting exponentially small
corrections similar to the equation above.

\subsection{Quantum field vs.~classical stochastic system}
Another reason why we need $\Lambda^{-1} \gg H^{-1}$ is that, in this approximation, as it was already noticed in the context of eternal inflation in  \cite{Linde:1986fd,Linde:1986fc}, we can neglect the quantum nature of the field and treat it as a classical stochastic system. Consider a single Fourier mode of the scalar field; as we are neglecting interactions, it simply behaves as an harmonic oscillator with parameters that depend on the background and are thus time dependent. 
As it is well known, in the limit in which the mode wavelength is much
larger than the Hubble scale, the system is in a squeezed state with
very large squeezing parameter \cite{Polarski:1995jg}. This
corresponds to the fact that the dynamics is dominated by the so
called ``growing mode", while the other independent solution dies away
exponentially (``decaying mode"). Although the system is still fully
quantum, we expect that all physical measurements and interactions
will only be sensitive to the growing mode so that the decaying one
can be disregarded. This leads to a classical system, with the
amplitude of the growing mode as a classical stochastic
variable \footnote{Referring to the discussion of footnote
  \ref{freivogel}, we can again assume that a period of non-eternal
  inflation, $\zeta \ll 1$, stretches all the relevant modes largely out
of the horizon, so that we are indeed allowed to treat them classically.}.

In conclusion, under our approximations the inflaton fluctuations are well described by a classical stochastic field with Gaussian statistics. All correlation functions are then determined uniquely by the two-point function. At equal times, before smoothing  this is
\be
 \langle \delta\phi(\vec x, t) \, \delta \phi (\vec y, t) \rangle = \frac{1}{(2 \pi)^2}\frac{1}{|\vec x- \vec y|^2 \, a^2(t)}
 - \frac{H^2}{(2\pi)^2} \, \log \frac{|\vec x - \vec y|}{L}  \; ,
 \ee
where $L$ is our IR cutoff---the comoving size of the universe. 
The first term is the usual two-point function for a massless scalar in Minkowski space. The second piece is peculiar to de Sitter space, and it is the celebrated scale-invariant spectrum.
Upon smoothing out the field, the above expression is the correct one at distances larger than the UV cutoff, $|\vec x - \vec y| \cdot e^{Ht} \gg \Lambda^{-1}$,
whereas at shorter distances the two points can be thought of as coincident and we get the classic result \cite{Vilenkin:1982wt,Linde:1982uu,Starobinsky:1982ee}
\be \label{variance}
 \langle (\delta\phi ^2 (\vec x, t) )  \rangle = 
 \alpha \Lambda^2 + \frac{H^2}{(2\pi)^2} \big(  \log \Lambda L + \beta + Ht \big)  = 
 \frac{H^3}{(2\pi)^2} \, t + {\rm const} \; .
\ee
As usual, the coefficient of the quadratically divergent piece, $\alpha$, as well as the finite part associated to the log-divergent piece, $\beta$, are order-one numbers that are regularization-dependent, i.e.~they depend on the precise filter-function we adopt to smooth-out the field. On the other hand, the coefficient in front of the log-divergence is regularization-independent, being determined by long-distance physics.
The factor of $t$ comes from the explicit time-dependence of the UV-cutoff in comoving coordinates, or, equivalently, from the time-dependence of the IR cutoff in physical coordinates. Physically, as time goes on more and more modes are included in the smoothed field and, given the scale-invariance of the spectrum, they all contribute to the variance (\ref{variance}). The coefficient of $t$ is also regularization-independent, being determined by the log-divergence coefficient and by how the cutoff scales with time---not by the precise definition of the cutoff.

We see that the variance of $\delta \phi$ at a given comoving point $\vec x$ increases linearly with time. This reminds us of a random walk and it is the base of the so called stochastic approach to inflation which dates back to \cite{starobinsky,Linde:1986fd,Linde:1986fc}. Indeed, if as smoothing procedure we choose a sharp cutoff in momentum space, the analogy becomes exact: at any given point $\delta \phi(\vec x)$ undergoes a random walk, every new mode included in the smoothing giving a random kick to the field, totally uncorrelated to the previous ones. This will allow us to exploit the technology of diffusion processes. 

Before moving to the actual calculations, we want to underline the
simplicity of the system under study. Everything has been reduced to a
free scalar in de Sitter space. Going towards the eternal inflation
regime however, we will see that this system gives rise to a kind of
phase transition producing an infinite reheating volume. Although we
are studying a free field, the relation between the reheating volume
and the scalar is very non-linear and this is what gives rise to the
interesting dynamics.

%%%%%%%%%%%%%%%%%%%%%%%%%%
%%				SERGEI					%%
%%%%%%%%%%%%%%%%%%%%%%%%%%

\section{Volume statistics\label{sec:volume}}

\subsection{The average}
\label{sec:average}
Let us study the probability density $\rho(V)$ for the volume
of the reheating surface $\phi = \phi_r$. Our goal is to identify a sharp transition in the behavior of this probability
density at some finite value of the parameters of the model. In the
setup discussed in the previous section the only dimensionless
parameter is given by the quantity $\dot \phi
/H^2$, the ratio between the classical motion and quantum fluctuations. We want to see
whether it is possible to identify a critical value of  $\dot \phi
/H^2$ where a sharp transition in the properties of $\rho(V)$
occurs. This would correspond to the onset of the eternal inflation regime.

Let us start with calculating the average reheating volume $\langle V\rangle$
\be
\label{defVav}
\la V\ra=\int dV\;V\rho(V)\;.
\ee
If we call $t_r(\vec x)$ the reheating time as a function of the
comoving coordinate $\vec x$, the average volume can be written as
\be
\langle V \rangle = \big \langle\int d^3x \; e^{3 H t_r(\vec x)}\big\rangle 
= \int d^3x \;  \big \langle e^{3 H t_r(\vec x)} \big\rangle \; ,
\ee
where we reversed the order of integration and averaging thanks to linearity of both.
Then the problem
becomes very simple:
\be
\label{Vavpr}
\la V\ra= L^3 \int \! dt \, e^{3Ht}p_r(t)\;,
\ee
where $p_r(t)$ is the probability that at a given point $\vec x$ the field reaches the reheating
value $\phi_r$ at time $t$. By translational invariance this probability does not depend
on the point $\vec x$, so that the integral over the comoving
coordinates can be factored out to give the volume of the comoving box $L^3$.

\begin{figure}
\includegraphics[width=8cm]{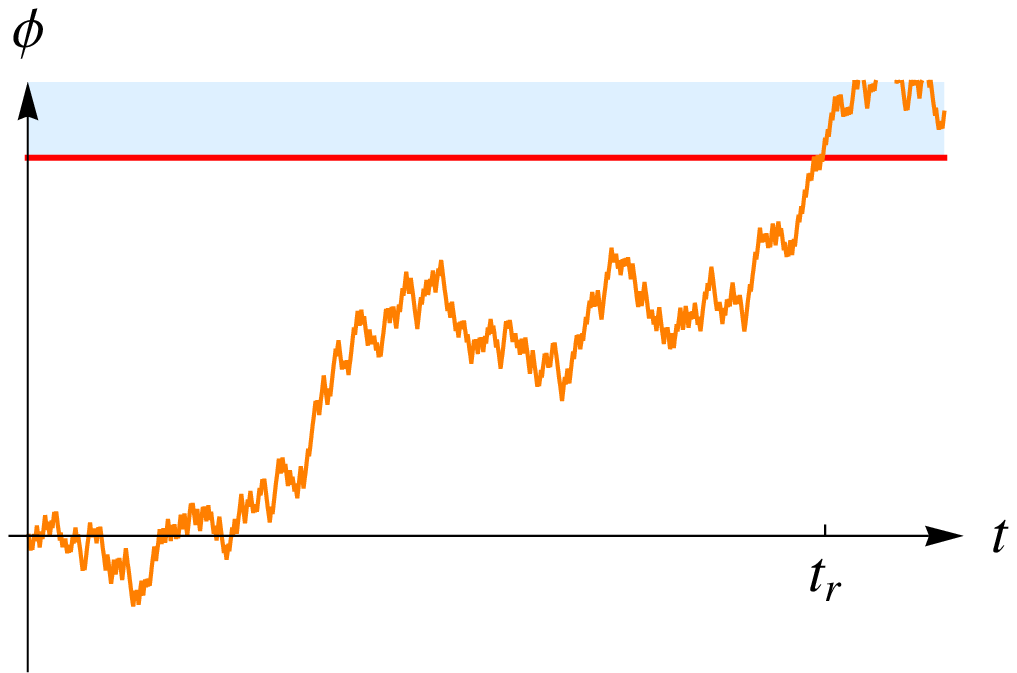}
\hfill
\includegraphics[width=8cm]{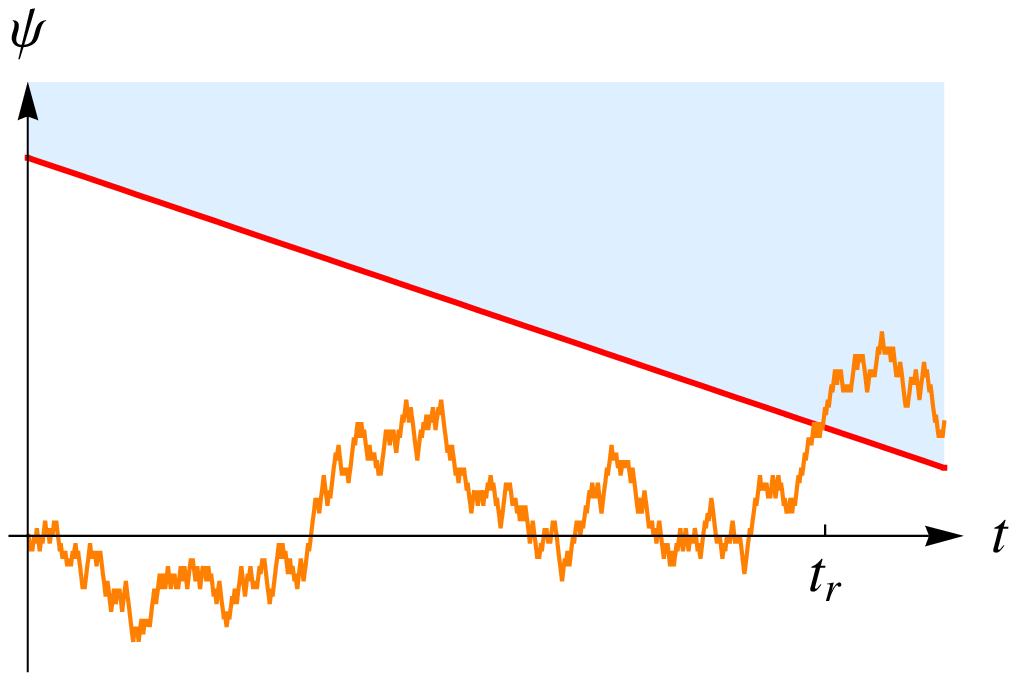}
\caption{\label{randomwalk} \small\it At any given comoving point, the local inflaton value undergoes a random walk. {\em Left:} for $\phi$ the reheating barrier is held fixed at $\phi = \phi_r$, but there is classical drift towards it.
{\em Right:} for the fluctuation $\psi$ there is no net drift, but the barrier is approaching at a speed $\dot \phi$.}
\end{figure}

To obtain the probability of reheating at time $t$ we have to start
from the probability $P(\phi, t)$ of having a given field value $\phi$ at time
$t$. This again does not depend on the point $\vec x$. 
Then the probability to reheat at time $t$ can be calculated as 
time-variation of the probability of still being inflating, {\em i.e} of having $\phi$ anywhere on the left of $\phi_r$:
\be
\label{reheating_prob}
p_r(t)= -\frac{d}{d t} \int^{\phi_r}_{-\infty} \! d\phi  \; P(\phi,t) \; .
\ee
As we already discussed in the previous section, at leading order in
the slow roll parameters,  the perturbation of the inflaton field around the classical slow roll 
solution,
\be
\label{phidef}
\psi=\phi-\dot\phi t\;
\ee
is a Gaussian field with variance $\sigma^2$ that grows linearly with
time (see figure \ref{randomwalk}),
\be
\label{sigma}
\sigma^2={H^3\over 4\pi^2}t \;.
\ee
Here we dropped $t$-independent terms present in the full expression (\ref{variance}). It is straightforward to keep track of these terms and check that they do not affect our results, that depend only on the late time asymptotics of $\sigma^2$. In fact the constant pieces in eq.~(\ref{variance}) can be set to zero by a constant shift of the time variable, $t \to t + {\rm const}$. All properties of the system that are dominated by the $t \to \infty$ limit---like whether inflation is eternal or not---are totally insensitive to such a shift. Note that the UV cutoff $\Lambda$ does not enter
in the $t$-dependent part (\ref{sigma}), nor does the IR cutoff $L$, so that our conclusions are not sensitive to the values of $\Lambda$ and $L$.
To avoid the proliferation of numerical coefficients, in what follows it is convenient to use $\sigma^2$ itself as time variable.
At any given $\vec x$, the probability distribution $P(\phi,t)$, when written as a function of $\psi$ and $\sigma^2$,
 satisfies the diffusion equation
\be
\label{diffusion}
{\d P\over\d \sigma^2}={1\over 2}{\d^2P\over \d\psi^2}\;.
\ee

Inflation ends when the inflaton field $\phi$ 
reaches $\phi_r$.
Consequently, we are interested in solving  the diffusion equation (\ref{diffusion})
with an ``absorbing barrier'' at the corresponding value of $\psi$, {\it i.e.}~the solution should satisfy the Dirichlet boundary condition
\be
\label{barrier}
P|_{\psi=\psi_r}=0\;,
\ee
with 
\be 
\psi_r=\phi_r - \dot\phi t = \phi_r - 4\pi^2{\dot\phi\over H^3}\sigma^2 \;.
\ee 
Such a solution can be constructed by using a generalization of the
method of images.
One starts with the standard solution describing diffusion without any
barriers and with initial condition localized at $\psi =0$ for $t = 0$
\be
\label{nobar}
P_0 (\psi, \sigma^2)= {1\over \sqrt{2\pi\sigma^2}}e^{-\psi^2/2\sigma^2}\;.
\ee 
In the case of a time independent barrier, it is well known that the Dirichlet boundary
condition can be satisfied adding a negative image located specularly
with respect to the barrier.
In our case we have a barrier moving linearly in time. It is
straightforward to show that in this case the solution becomes
\be
\label{image}
P_1(\psi, \sigma^2)={1\over \sqrt{2\pi\sigma^2}}\l e^{-\psi^2/2\sigma^2}- e^{8 \pi^2 \dot\phi\, \phi_r/H^3} e^{-(\psi- 2 \phi_r)^2/2\sigma^2}\r\;.
\ee  
The image is symmetric with respect to the barrier position at time $t=0$ and it is multiplied by a constant factor which depends on the barrier speed. 
 Note that in the whole physical region, $\psi < \psi_r$, the contribution from the image is 
 smaller than the $P_0$ term, as it should be in order for
 the density distribution $P_1$ to remain positive. Far from the barrier the contribution of the image is exponentially suppressed.

We can now use our solution for $P$ to evaluate the probability (\ref{reheating_prob}) of reheating at a given time $t$. By making use of the diffusion equation (\ref{diffusion}) and taking into account that $P$ vanishes on the barrier, eq.~(\ref{barrier}), we have
\be
p_r(t) = -\frac{H^3}{8 \pi^2} \left.\frac{\partial P}{\partial \phi}\right|_{\psi=\phi_r-\dot\phi t} \;.
\ee
Disregarding the polynomial dependence in front of the exponential we have
\be
\label{eq:prexp}
p_r(t) \sim e^{-2 \pi^2 \frac{(\phi_r-\dot\phi t)^2}{H^3 t }} \;.
\ee
At late times this gives for the volume (\ref{Vavpr})
 \be
 \label{Vavlate}
 \langle V\rangle \simeq L^3 \int_0^\infty dt\;\alpha(t) \, e^{3Ht-{2\pi^2\dot\phi^2\over H^3}t+O(1)}\;,
 \ee 
 where the prefactor $\alpha(t)$ does not have exponential dependence on time.
 We see that the average reheating volume remains finite if and only if
 \be
 \label{voldiv}
 \Omega\equiv{2\pi^2\over 3}{\dot\phi^2\over H^4}>1\;.
 \ee
 So far we have not talked about the initial condition we choose for
 the inflaton. Clearly, as the general solution of the diffusion
 equation is given by the convolution of $P_1(\psi,t)$ with the
 initial probability distribution, the initial condition is irrelevant
 for the convergence of the average reheating volume.
 A similar criterium for the divergence of the inflating volume was
 put forward in \cite{Winitzki:2001np}. Here we focus on the more physical quantity of
 reheating volume and we proceed to better characterize what happens
 at $\Omega =1$.

Equation (\ref{voldiv}) appears as a sharp criterion for whether eternal inflation takes place or not.
At the end we will argue that this is essentially correct; however to justify this conclusion we first need to  address several important subtle points.
Let us try to understand in more detail what happens to the probability distribution for the reheating
 volume $\rho(V)$ at the critical value  $\Omega=1$, where the average reheating  volume
diverges.
There are two possible options. First, it may happen that each moment 
$\la V^n\ra$ of the distribution has a different critical value of $\Omega$ when it starts diverging.
This would happen, for instance, if $\rho(V)$ had a power law asymptotics at large values of $V$ with
the power depending on $\Omega$, e.g.
\be
\label{powerrho}
\rho(V)\propto {1\over 1 + V^{1+\Omega}} \;.
\ee
If this were the case, the value $\Omega=1$ associated to the divergence of $\langle V \rangle$ would not really mark any sharp transition in the
physical behavior. Each moment of the distribution would single out a different transition point. Perhaps in this case a better characterization of the onset of eternal inflation would
be given by the value of $\Omega$ where the probability distribution itself ceases to be normalizable
($\Omega=0$ for the model distribution (\ref{powerrho})), if such a value exists. 

The other possibility is that at $\Omega=1$ all moments of the probability distribution becomes infinite, including
the ``zeroth" moment, i.e.~the probability distribution also ceases to be normalizable at this point.
This would happen, for instance, for a family of probability distributions of the form
\be
\label{exprho}
\rho(V)\propto q(V)e^{(1-\Omega)V}\;,
\ee
 where $q(V)$ has a power-law behavior at large $V$. In this case the critical value $\Omega=1$ indeed has a very sharp physical
 meaning. 
 
 We will see that the actual answer is somewhat more complicated than both of these options. However, already at this point it is possible to argue that in the setup we are considering  the distribution is not of the type (\ref{exprho}) and that different moments diverge at different values of $\Omega$. Indeed, it is straightforward to see that for 
 {\it any} given value of $\Omega$ there is a value $n_0$ such that all moments of the volume distribution $\la V^n\ra$ with $n>n_0$ diverge.  Indeed, let us imagine that  $\Omega$ is much larger than one, so that naively one is very far from the eternal inflation regime.
Let us consider a class of inflaton trajectories that at time $t$ in one Hubble patch have a huge quantum fluctuation in the direction opposite to the classical motion, such that the inflaton ends at
 \be
 \phi=\phi_r-\dot\phi t\;.
 \ee
The classical trajectory would have $\phi = \dot\phi t$, so that at large $t$
the probability of such a fluctuation is exponentially suppressed by a factor of order $e^{-A\cdot H t}$,
with $A\sim\dot\phi^2/H^4 \sim \Omega^2$. The important point is that the exponent grows only linearly with time (and not quadratically).
The reason is that  the variance $\sigma^2$ of the inflaton perturbation $\psi$  is itself a linear function of $t$.
Then, with probability close to one such a trajectory will give rise to a reheated volume at least as
big as $e^{3Ht} H^{-3}$, since this is the volume one would get just classically rolling down from $\phi=\phi_r-\dot\phi t$ to $\phi = \phi_r$. Consequently, for $n>A/3 \sim \Omega^2$ the contribution to the moment $\la V^n\ra$ from this set of trajectories grows exponentially as a function of $t$---the large power of the exponentially growing volume over-compensates for the small probability of the required fluctuation---and these moments diverge. 

At this point one is tempted to conclude that the situation is analogous to what happens
with the distribution (\ref{powerrho}) and that there is no sharp transition to the eternal inflation regime.
We will argue later that this conclusion is too quick, and that instead there is a sharp transition. Notice that for the rough argument about higher moments to work, it is essential to consider, as time grows, backward fluctuations that go further and further along the potential without bound. 
To see what this implies,  it is instructive to understand the effect we just described
at a more quantitative and detailed level. For this purpose we are going to study now the behavior of $\la V^2\ra$ and show that indeed it starts diverging at a critical value $\Omega > 1$, {\em i.e.} when the average volume $\langle V \rangle$ is still finite.

%%%%%%%%%%%%%%%%%%%%%
\subsection{The variance}
\label{sec:variance}

The calculation we did for the average is deceivingly simple as the
only thing we needed was the 1-point probability $P(\phi,t)$ to have a
value of the field $\phi$ at time $t$, which is clearly independent of
the comoving position $\vec x$. However the study of the full
probability for the reheating volume $\rho(V)$ is not straightforward
as the only expression for it is based on a functional integration
over the field realizations 
\be
\label{fi}
\rho(V)=\int \! {\cal D}\phi \: {\cal P}[\phi] \: \delta \big( V- \textstyle{\int} \!d^3x \, e^{3Ht_r(\vec x)} \big) \;,
\ee
where ${\cal D}\phi$ is the functional  measure on the set of all possible space-time realizations
of the inflaton field, ${\cal P}[\phi]$ is the probability of a specific realization and $t_r(\vec x)$ is the reheating
time for a given realization as a function of the comoving coordinate
$\vec x$. 
% To make sure that this result agrees with the formal expressions (\ref{defVav}), (\ref{fi}) above, note that by inserting
% \[
% 1=\int dt\; \delta(t_r(x)-t)
% \]
% into the functional integral (\ref{fi}) one arrives at the same result (\ref{Vavpr}) with
% \[
% p_r(t)=\int{\cal D}\phi{\cal P}[\phi]\delta(t_r(x)-t)\;.
% \]

Instead of trying to evaluate the expression above, to calculate $\la
V^2\ra$ we will follow a procedure similar to the one used for the average.
Namely, let us define the 2-point probability distribution of the
inflaton field $P_2(\phi_1,\phi_2, | \vec x_1- \vec x_2 |,t_1,t_2)$, such that
\be
P_2 \, d\phi_1 d\phi_2
%dt_1dt_2dx
\ee
gives the joint probability of finding the inflaton field in the ranges $(\phi_{1(2)},\phi_{1(2)}+d\phi_{1(2)})$
at the two space-time points $(\vec x_1, t_1)$ and $(\vec x_2, t_2)$. For
translational and rotational invariance this joint probability just
depends on the distance between the two points $| \vec x_1- \vec x_2 |$. Analogously to the previous section we can define a probability density $p_{2r}(t_1,t_2, |\vec x_1- \vec x_2|)$ for  reheating to happen
at the comoving point $\vec x_1$ at time $t_1$ and at the comoving
point $\vec x_2$ at time $t_2$. Similarly to (\ref{reheating_prob}) this will be related to $P_2$ by 
\be
\label{reheating_prob2}
p_{2r}(t_1,t_2,|\vec x_1- \vec x_2|)=-
\d_{t_1}\d_{t_2}\int^{\phi_r}_{-\infty} \!
d\phi_1\int^{\phi_r}_{-\infty} \! d\phi_2\; P_2(\phi_1,\phi_2, |\vec
x_1- \vec x_2|,t_1,t_2) \;.
\ee
%in the time intervals $(t_{1(2)},t_{1(2)}+dt_{1(2)})$ for two points separated by a distance in the range
%$(x_1-x_2,x_1-x_2+dx)$. Then, similarly to (\ref{Vav}) one has
Then the expectation value of $V^2$ can be written as 
\be
\label{VVavpr}
\la V^2\ra= \left\langle \int d^3x \,d^3y \,e^{3Ht_r(\vec x)}\, e^{3Ht_r(\vec y)}\right\rangle = L^3 \int \! dt_1 dt_2 d^3x \;e^{3Ht_1} e^{3Ht_2} \; p_{2r}(t_1,t_2, |\vec x|)\;.
\ee

To study the convergence of this integral we need to find the 2-point
probability $P_2$.  For fixed comoving distance $|\vec x|$ between the two points this
probability has a very different behaviour at early and late times.  
At early times, 
\be
\label{earlyt}
t_1, t_2 \ll t_*\equiv -{H^{-1}}{\log H |\vec x|}\;,
\ee
the two points are within one Hubble patch and therefore they are not
resolved by the smoothing procedure. As such their evolution is
perfectly correlated. In particular the 2-point distribution at equal
times $t_1 = t_2$ is just given by the 1-point 
distribution (\ref{image}),
\be
\label{earlyP}
P_2(\psi_1,\psi_2, |\vec x|,t,t) \simeq P_1(\psi_1,t)\delta(\psi_1-\psi_2)
\quad \mbox{ for } \quad t \ll t_*(|\vec x|)\;.
\ee
In the opposite limit, when the distance is much larger than the
Hubble radius, the inflaton fluctuations at the two points are
completely uncorrelated. 
In this case, the probability distribution $P_2$ satisfies a diffusion equation with respect to 
each pair of variables $(t_1,\phi_1)$, $(t_2,\phi_2)$,
\be
\label{diffusion12}
{\d P\over\d \sigma_{1(2)}^2}={1\over 2}{\d^2P\over \d\psi_{1(2)}^2}
\quad \mbox{ for } \quad t_1, t_2 \gg t_*\;.
\ee
In the equations above $\psi_{1(2)}$ and $\sigma^2_{1(2)}$ are related to $t_{1(2)}$ and
$\phi_{1(2)}$ by the relations (\ref{phidef}) and (\ref{sigma}).

\begin{figure}
\begin{center}
\includegraphics[width=11cm]{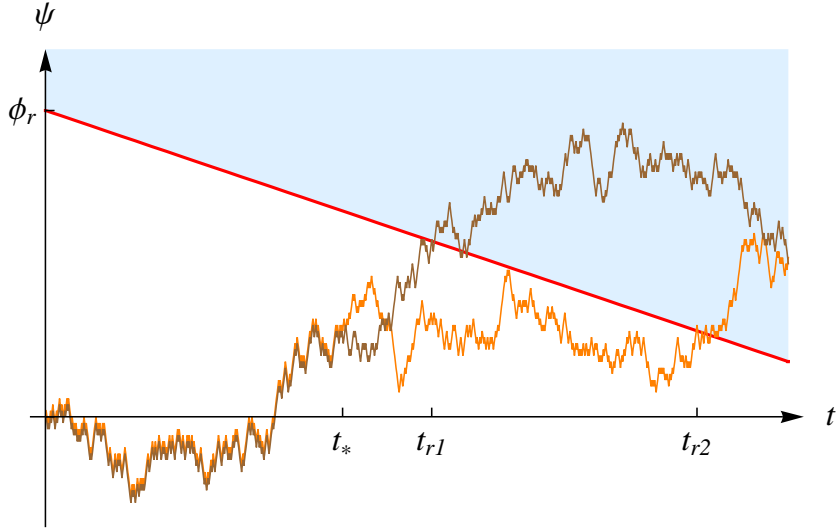}
\caption{\label{randomwalk2} \small \it  Our two-step approximation. The local inflaton fluctuations in two nearby comoving points are assumed to be totally correlated until the physical distance between the two points exits the UV cutoff  ($t<t_*$), and totally uncorrelated afterwards ($t>t_*$).}
\end{center}
\end{figure}

In the following we are going to use a ``two steps'' approximation in
which we consider a single random walk before $t_*$ and two
uncorrelated ones after $t_*$, as dipicted in figure \ref{randomwalk2} (\footnote{The same approximation has
  been used in the study of structure formation in
  \cite{Scannapieco:2002qd}.}). 
The exact solution is more complicated
as there is an intermediate period of partial correlation \cite{Scannapieco:2002qd}. However it
will be easy to check that our results will not depend on the
exact definition of $t_*$, when the two regimes are matched. This
strongly indicates that taking into account the period of partial
correlation would not change the range of $\Omega$ when
$\langle V^2 \rangle$ converges.

In this 2-step approximation, the 2-point probability when the distance between the two points is
larger than Hubble is obtained first following a single random walk until
the time of ``separation'' $t_*$ and then convolving this initial condition with
two independent Green's functions for the diffusion equation with a
barrier. This translates into
\be
\label{convolution}
P_2(\psi_1,\psi_2, |\vec x|,t_1,t_2)=\int_{-\infty}^{\phi_r - \dot\phi t_*} \!\!\!\! d\psi\, G(\psi_1,\psi,t_1-t_*)G(\psi_2,\psi,t_2-t_*)P_1(\psi,t_*)\;,
\ee
with $t_1, t_2 > t_*$. Notice that the dependence on $|\vec x|$ is hidden in $t_*$, given by equation (\ref{earlyt}).
The Green's function $G$ is straightforward to construct by making use of the solution (\ref{image}). Namely, it is equal
to
\be
\label{green}
G(\psi_1,\psi,t)=\tilde{P}_1(\psi_1-\psi,t-t_*)
\ee
where ``tilde" indicates that $\phi_r$ in eq.~(\ref{image}) must be replaced by
\be
\phi_r - \psi -\dot\phi t_*
\ee
to take into account that the independent motion of each of the two points starts at $\psi$ at time $t_*$, rather than at the origin at $t=0$.
%\be
%\tilde\psi_r=\psi_r(t-t_*)-\psi\;.
%\ee

To prove that the variance diverges when the average volume is still
finite, it is enough to obtain a lower bound on the integral
(\ref{VVavpr}). As the integrand is positive definite, we can
therefore restrict the range of integration by considering only pairs
of points $(\vec x_1, \vec x_2)$ in different Hubble patches (both at
$t_1$ and at $t_2$). In other words we can impose a UV cutoff on the integration over the comoving
volume in eq.~(\ref{VVavpr}) by requiring
\be
\label{UVcut}
t_*(|\vec x|)< t_1,t_2 \;.
\ee
It is anyway straightforward to check that the remaining range of
integration gives a contribution that always converges whenever $\Omega >1$. 

As before we can use (\ref{diffusion12}) to evaluate the integrals in (\ref{reheating_prob2}) and arrive at the following expression
\be
\label{VVav}
\langle V^2\rangle=-{H^6 \cdot L^3\over 64\pi^4}\int_0^\infty
dt_1dt_2\int_{|\vec x|>H^{-1}e^{-H\tau}} \!\!\!\!\!\!\!d^3x\;e^{3H(t_1+t_2)}\d_{\psi_1} \d_{\psi_2}P_2(\psi_1,\psi_2, |\vec x|,t_1,t_2)\Big{|}_{\psi_{1(2)}=\phi_r-\dot\phi t_{1(2)}},
\ee
where $\tau = {\rm min}(t_1,t_2)$.

The partial derivatives act on each of the two Green functions of eq.~(\ref{convolution}) and give a dependence analogous to the 1-point case eq.~(\ref{eq:prexp}). The evolution before splitting given by $P_1$ contains the image term and one should sum the two contributions. However, as already discussed in the previous section, we will find that the divergence of $\langle V^2 \rangle$ is dominated by trajectories which are exponentially far from the barrier as time becomes large. This implies {\em a posteriori} that the image term is completely irrelevant.

The convolution integral in (\ref{convolution}) can be done in saddle point approximation. The saddle point is given by
\be
\label{psisaddle}
\psi_{\rm saddle} =t_*
{
\l \psi_1+\psi_2\r
t_*-t_2\psi_1-t_1\psi_2\over t_*^2-t_1t_2 
} \;.
\ee
This gives for the exponential part of $P_2$
\be
\label{P2saddle}
%P_2(\psi_1,\psi_2,|\vec x|,t_1,t_2)\sim \exp{\l-3H\Omega{t_2\psi_1^2-2t_*\psi_1\psi_2+
%t_1\psi_2^2\over t_1t_2-t_*^2}\r}\;.
P_2(\psi_1,\psi_2,|\vec x|,t_1,t_2)\sim
\exp{\l-\frac{2\pi^2}{H^3}\cdot {t_2\psi_1^2-2t_*\psi_1\psi_2+
t_1\psi_2^2\over t_1t_2-t_*^2}\r}\;.
\ee
Note that this expression non-trivially depends on the spatial separation $|\vec x|$ through the horizon crossing time $t_*$, as shown in (\ref{earlyt}). To plug this result back in (\ref{VVav}) it is convenient 
to use 
\be
t_\pm=t_1\pm t_2
\ee
as time variables and the horizon exit time $t_*(|\vec x|)$ in place of the comoving separation $\vec x$.
With this change of variable and keeping only the leading exponential behavior at large times as in the previous section, one obtains (restricting to $t_1 \geq t_2$)
\be
\label{saddleV2}
\la V^2\ra\sim\int_0^\infty dt_+ \, e^{3Ht_+}\int_0^{t_+}dt_-\int_0^{t_+-t_-\over2} \!\!\!\! dt_* \,e^{-f(t_+,t_-,t_*)}\;,
\ee
where
\be
\label{exp}
f(t_+,t_-,t_*)=3Ht_*+3H\Omega\l t_+-2t_*+{4t_*^2(t_+-2t_*)\over t_+^2-t_-^2-4t_*^2}\r\;.
\ee
The first term on the right hand side of this equation comes from the change of integration measure from $x$ to $t_*(|\vec x|)$.
To check whether the integral over $t_+$ converges, we can minimize the function $f$ with respect to $t_-$ and $t_*$ in the range of integration.  As $t_-$ enters into $f$ only in the denominator of the last term, it is clear that the saddle point is 
 \be
 t_-=0\;.
 \ee
 Then the minimization with respect to $t_*$ gives
 \be
 \label{t*min}
 t_{*min}={t_+\over 2}\l\sqrt{2\Omega}-1\r\;,
 \ee
 and the value of $f$ at this minimum is
 \be
 \label{fmin}
 f(t_+,0,t_{*min})=3Ht_+\l\sqrt{2\Omega}-{1\over 2}\r\;.
 \ee
 This grows faster than $3Ht_+$ for 
 \be
 \label{epscr}
 \Omega>{9\over 8}\;.
 \ee
 We conclude that $\la V^2\ra$ ceases to be finite when the parameter 
  $\Omega$ drops below $9/8$, {\it i.e.}~before the average volume $\la V\ra$ starts diverging.
  As a consistency check of the calculation, note that at $\Omega=9/8$ the optimal horizon exit time
  $t_{*min}$ indeed belongs to the integration interval in (\ref{saddleV2}). For $\Omega>2$
  this would no longer be true and the dominant contribution would come from $t_*=t_+/2$: $\langle V^2 \rangle$ would be dominated by the dynamics before the two points become independent.

%%%%%%%%%%%%%%%%%%%%%%%
\subsection{Uphill barrier}
\label{sec:second}
The above explicit calculation of $\la V^2\ra$ nicely fits into the general argument presented at the end of section \ref{sec:average}. We have seen that, in saddle point approximation, the dominant contribution to $\la V^2\ra$ comes from points that reheat at the same time $t_1 = t_2$. These points start evolving independently when separated by a physical distance $\sim H^{-1}$; this happens at a fraction of the eventual reheating time, which is equal to $1/2$ for the critical value $\Omega = 9/8$, see (\ref{t*min}).  

Where is the most likely position along the potential at $t_*$, when the two points start their independent evolution? From the saddle point result (\ref{psisaddle}) we see that it is a position with a huge backward  fluctuation with respect to the classical trajectory
\be
\label{psimin}
\psi=-\sqrt{2\over \Omega}\dot\phi \, t_{*min}\;.
\ee
For $\Omega<2$ this fluctuation corresponds to a negative value of the
inflaton 
\be
\label{eq:saddleback}
\phi=\left(1-\sqrt{2/\Omega}\right) \cdot \dot\phi t_{*min}\,,
\ee 
so that the most likely position at $t_{*min}$ is higher than the
starting point at $t=0$ and it becomes higher and higher without bound
for $t_{*min} \to \infty$. This agrees with the general argument above
\footnote{Notice that as $\langle V^2\rangle$ is dominated by a splitting point infinitely far away from the barrier it is justified disregarding the image terms of the probability $P_1$ in (\ref{convolution}).}.

To summarize, all this sounds as if  the 
volume distribution has qualitative properties similar to the model distribution (\ref{powerrho}), where the various moments start diverging at different values of $\Omega$. Thus it seems that the value 
$\Omega=1$, where the average volume diverges, does not really correspond to any sharp physical
transition. However this conclusion is too quick. In section 
\ref{sec:bacteria}, where we study a discretized version of the same system, we will argue that at $\Omega=1$ another sharp transition occurs besides the divergence of $\langle V \rangle$: the total probability of having a finite reheating volume becomes smaller than one! This property is expected in the transition into the eternal inflation regime as it tells us that there is a finite probability of an infinite reheating volume or equivalently for inflation to last forever. Clearly this sharply defines $\Omega =1$ as the transition point to eternal inflation. Before moving to this issue, let us conclude our study of the moments of the distribution $\rho(V)$.

%This property is not present in either of the naive
%model probability distributions (\ref{powerrho}), (\ref{exprho}), but a very natural one to expect at the
%transition to the eternal inflation regime. It is simply saying, that there is a finite probability to
%have an infinite reheated volume, {\it i.e.} for inflation to last forever!

As we stressed, the divergence of the higher moments is qualitatively
different with respect to the case of the average. The explicit
calculation of $\langle V^2 \rangle$, following the general argument
made in section \ref{sec:average}, shows that the divergence is related to the possibility of fluctuating uphill without bound, infinitely far from the barrier. What we are going to check is that, if we put a bound to the possibility of going infinitely uphill, {\em all the moments diverge at the same critical value $\Omega =1$.}

To impose a limit on fluctuations infinitely far from the reheating
point, we can put a reflecting barrier so that the field value is
constrained to satisfy
\be
\label{UVlimit}
\phi > \phi_{up}\;,
\ee
where $\phi_{up}$ will be a large negative value. The value of $\psi$
at the barrier will be given by
\be
\label{psibUV}
\psi_{up}=\phi_{up}-\dot\phi t\;.
\ee
 
% This implies, that the diffusion equations for the distributions of inflaton perturbations $\psi$  acquire
%  a second barrier,
%  \be
% \label{UVbarrier}
% P|_{\psi=\psi_{UV}}=0\;,
% \ee
% where the value of $\psi$ at the barrier is equal to

Here we are going to study the effect of this new barrier on the
calculation of $\langle V^2 \rangle$, while we postpone the discussion
about all higher moments to Appendix \ref{app:moments}.
We do not need to find the explicit form of the solution of the
diffusion equation with two barriers.
To keep track of the late time exponential behavior for large enough 
$|\phi_{up}|$, it is enough to consider that the integration range in
the convolution (\ref{convolution}) is now restricted to
$(\psi_{up},\psi_r)$ and it is not extended infinitely uphill. 
For $\Omega<2$ the saddle point position (\ref{eq:saddleback}) moves
backwards and thus it eventually hits the new uphill barrier. 
The saddle point exits the physical region and therefore, in the
leading exponential approximation, the result will be dominated by a
splitting point position on the uphill barrier, {\em i.e.} the point
closest to the (would-be) saddle.
In this approximation one obtains for the 2-point distribution
\be
\label{P2bar}
P_2(\psi_1,\psi_2,|\vec x|,t_1,t_2)\simeq 
\exp{
\left[
- \frac{ 2 \pi^2}{H^3}
         \l
               {\l \psi_1-\psi_{up} \r^2\over t_1-t_*}
+
                {\l\psi_2-\psi_{up}\r^2\over t_2-t_*}
+
                {\psi_{up}^2\over t_*}
          \r
\right]}\;.
\ee
We are interested in the late time behaviour of this expression. The
position of the uphill barrier $\phi_{up}$ becomes irrelevant in this limit
as it is dominated in (\ref{psibUV}) by the time dependent
term. This means that $\phi_{up}$ disappears from the calculations and
therefore from the final critical value of $\Omega$ and
it can be thought just as a regulating device, which can be sent to
infinity. Notice however that its effect is crucial. Mathematically this comes
from the fact that the large $\phi_{up}$ limit and the late time limit
do not commute, because the saddle point (\ref{psisaddle}) eventually 
exits the integration region at any finite value of
$\phi_{up}$. In more physical terms this means that no matter how far
the barrier is there will be a sufficient large time such that 
\be
H^3 t \gtrsim \phi_{up}^2
\ee
so that the diffusion is so large to be sensitive to the presence of
the extra barrier. As the transition to eternal inflation is sensitive
to arbitrarily large times, the presence of the regulating barrier
will be important, independently on how far we put it.

To see that the presence of the second barrier changes the value at which
$\la V^2\ra$ diverges, let us plug (\ref{P2bar}) in the expression
(\ref{VVav}) for the $\la V^2\ra$. We obtain the same result as (\ref{saddleV2}) with the function $f$ now taking the form
\be
\label{expb}
f(t_+,t_-,t_*)=3Ht_*+3H\Omega\l t_+-t_*\r
\ee
We see that now the integral for $\langle V^2 \rangle$ converges for $\Omega >1$. Note that the
integral over $t_*$ is saturated at the upper limit, justifying
a posteriori an assumption implicit in the whole derivation---that in
the late time limit (meaning large $t_{1,2}$ limit) $t_*$ also
indefinitely grows.  

On the other hand it is easy to check that there is no effect of the
second barrier on the calculation of $\langle V \rangle$; this is
just a consequence of the fact that solutions to the diffusion
equation behave continuously  when one removes one of the barriers to
infinity. This is checked in Appendix \ref{app:moments}, where we study the explicit solution
in the presence of two barriers.

In conclusion, in the presence of an uphill regulating barrier,
$\Omega=1$ is the critical point both for $\langle V \rangle$ and for
$\langle V^2 \rangle$. In Appendix \ref{app:moments} we will see
that this holds also for all higher moments $\langle V^n \rangle$.

%%%%%%%%%%%%%%%%%%%%%%%%%%
%%				LEONARDO			%%
%%%%%%%%%%%%%%%%%%%%%%%%%%

\section{Bacteria model of inflation\label{sec:bacteria}}

The above results strongly suggest that $\Omega=1$ sets the transition to
eternal inflation. To conclusively confirm this one would like to directly check that at 
$\Omega<1$ there is a finite probability for inflation to run forever. A natural quantity
to look at to address this issue is the total probability to have a finite reheating volume,
\be
\label{order_parameter}
P_f=\int dV\rho(V)\;,
\ee
where $\rho(V)$ is the volume probability distribution defined by eq.~(\ref{fi}). Clearly, $P_f=1$ if
inflation always ends in a finite time. On the other hand,  as we will see in this section, for $\Omega<1$ the total probability $P_f$  drops below one. This is a direct signal of eternal inflation---in this regime there is a finite probability $(1-P_f)$ that the total volume of the reheating surface is infinite. This result also shows
that the transition to the eternal regime is quite similar to a second order phase transition, with
$(1-P_f)$ being an order parameter. 

To derive this result one needs more detailed information about the probability distribution
$\rho(V)$ rather than just its multipole moments. This seems to be hard to achieve, as the only
available definition of $\rho(V)$ is a rather formal functional integral formulae (\ref{fi}).
As usual with functional integrals, to gain more control over $\rho(V)$ it is natural to switch to a discretized description
of the inflationary dynamics. This will be the approach of the current section.

%We now switch to a discrete analogue of our system. This will allow us to gain additional analytical control of our results. In particular we will consider the case in which both time evolution and continuous field values are replaced by discrete steps.

\begin{figure}[th!!]
\begin{center}
\includegraphics[width=12cm]{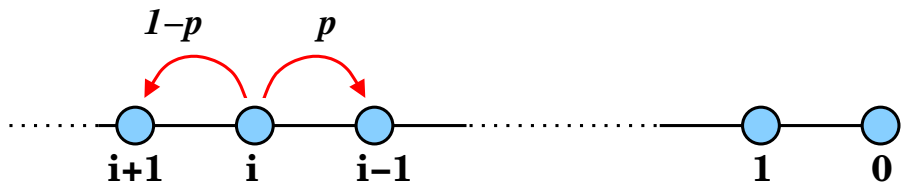}
\caption{\label{fig:branching process} \small \it The branching process.}
\end{center}
\end{figure}

With a biological analogy, consider at $t=0$ a bacterium that can live in a discrete set of positions along a line (see fig.~\ref{fig:branching process}). At $t=1$ the bacterium splits into $N$ copies. Then, each bacterium (independently of all the others) hops with probability $p$ to the neighboring site on its right, and with probability $(1-p)$ on the left.  $N$ and $p$ are fixed numbers. At $t=2$ each second-generation bacterium reproduces itself, and so on. The analogy with our inflationary system is clear: each bacterium represents an Hubble patch; sites are inflaton values. Reproduction is the analogue of the Hubble expansion; at  every $e$-folding $\sim e^3$ new Hubble volumes are produced starting from one. From then on the inflaton inside each Hubble volume evolves independently, with a combination of classical rolling and quantum diffusion. This is represented by
the random hopping of our bacteria. The difference in the probabilities of moving right and left gives a net drift, and thus corresponds to the classical motion. To complete the analogy we have to assume that there is a ``reheating'' site, $i=0$ in the figure: when a bacterium ends up there it stops reproducing and moving around---it dies.
In the previous sections we studied the statistical properties of the reheating volume $V$ as a function of $\Omega$. In the bacteriological analogy the reheating volume corresponds to the number of dead bacteria (= non-reproducing Hubble patches) in the asymptotic future. For analogy we denote the latter quantity by $V$, which of course now takes discrete values. Our task is to study the probability distribution of $V$ as a function of the parameter $p$.
A discrete system like the one we described goes under name of {\em branching process}, more precisely a multi-type Galton-Watson process (see e.g.~ref.~\cite{branching_books}).
Similar models have been studied in the context of eternal inflation in
\cite{Winitzki:2005ya}. Here the focus will be the relationship with the
continuum case and the precise characterization of what happens at the
critical point.

Before moving to the explicit study of the model,
let's summarize the main results we will achieve:
\begin{itemize}
\item We will explicitly show that in a suitable continuum limit the discrete model precisely converges to the full inflationary case we studied in the previous sections.  

\item There is a critical value
$p_c>1/2$ such that for $p \le p_c$ the {\it expectation value} $\la V \ra$ of
the number of dead bacteria at future infinity  diverges, whereas it converges for larger $p$'s. For inflation, $p_c$ is
the analogue of the critical $\Omega$ below which the expectation value of
the reheating volume becomes infinite. We shall see that in the continuum limit 
 the value of the critical $\Omega$ that we can infer from the discrete model matches the one
 obtained directly in the inflationary case.

\item The behavior of the variance of $V$ crucially depends on whether we consider a semi-infinite or finite line of sites.
In the semi-infinite case the variance starts diverging at a new critical value for $p$ that is larger than the above $p_c$ and agrees with the corresponding value of $\Omega$, eq.  (\ref{epscr}).
In the finite case on the other hand, the two critical values coincide, and the same happens for higher moments as well. 
This perfectly matches what we found in the inflationary system, where the critical behavior of the variance and higher moments
is sensitive to whether arbitrarily large field excursions are allowed.

\item The most dramatic way of characterizing the transition at $p = p_c$  is that {\em  the probability distribution develops a finite
probability for strictly infinite $V$}.
 In other words, there is a finite probability for the branching process not to terminate. 
This indicates that also in the inflationary case, for $\Omega < 1$ there is a finite probability of having an infinite reheating volume.
This is the most  pristine definition of eternal inflation---a finite probability of never ending inflation globally.

Notice however that there is always a finite probability to end inflation everywhere: for instance if in all Hubble patches  the inflaton always fluctuates towards reheating, we have a global exit from inflation. The probability for this to happen is clearly non-zero, but it becomes smaller and smaller the larger the initial inflationary volume. In the limit of a very large initial volume the transition at $\Omega = 1 $ becomes completely sharp: the probability of an infinite reheating volume jumps from zero to one.

To get further intuition about how a finite probability for infinite $V$ develops we will study
the distribution of dead bacteria as a function of time in the simplest case of just two sites.

\end{itemize}

\subsection{Recovering the inflaton dynamics from the discrete model}
The bacteria branching process described above clearly captures many of the qualitative
features of the actual inflaton dynamics, but it is not totally obvious that it allows to take the 
continuum limit that {\it exactly}
reproduces the inflationary results. To explain how this happens, let us start by describing a
different discrete model, which has a straightforward continuum limit reproducing inflation. This model
is harder to analyze than the bacteria process, but we will explain why the two are equivalent for our purposes.

To begin with, note that in the branching process description we have two different discretizations. 
First we have a discrete time step; for $N\sim e^3$ one step of the branching process corresponds
to a time interval of order $H^{-1}$. Second, the possible values of the inflaton field are discretized as
well, and labeled by the bacteria positions. However, to reproduce the inflationary results
we do not actually need to get rid of the time discretization. Indeed, as discussed in 
section~\ref{sec:smoothing}, a space-time smoothing of the scalar field over the scale $\Lambda^{-1}\gtrsim H^{-1}$ is understood in all our calculations. The physical results, such as the transition point,
do not depend on the details of the smoothing procedure and an explicit space-time discretization
(as in the branching process approach) is as good as the smooth filter (\ref{eq:smooth}). Hence we only need
to take the continuum limit in field space to reproduce the continuum results.

In the ``two steps" approximation we have been using in studying inflation---see the discussion below eq.~({\ref{diffusion12})---the inflaton dynamics is fully
determined by the diffusion equation for the probability of the inflaton values at one point in spacetime,
or equivalently, by the gaussian distribution (\ref{nobar}) for the inflaton fluctuations. A straightforward
way to approximate this dynamics by a branching process is to consider the following sequence of 
processes. For the $k$-th process of the sequence at each step every bacterium splits into $\tilde N$ copies. Then each
bacterium hops with probabilities $p_i$ ($i=-k,\dots, k$) to the $2k$ adjacent sites (see Fig.~{\ref{gaussian}). 
\begin{figure}[t!]
\begin{center}
\parbox{10cm}{\includegraphics[width=10cm]{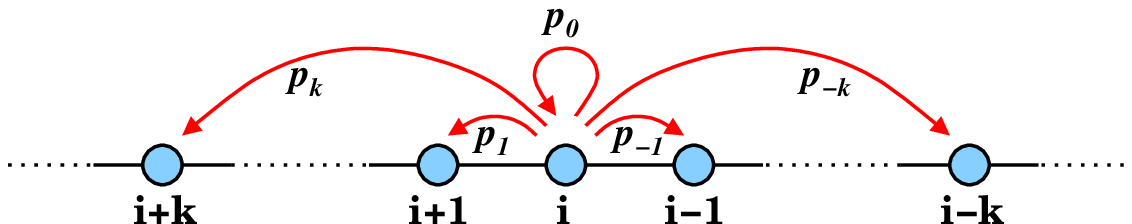}}
\hspace{.5cm}
\parbox{6cm}{\includegraphics[width=6cm]{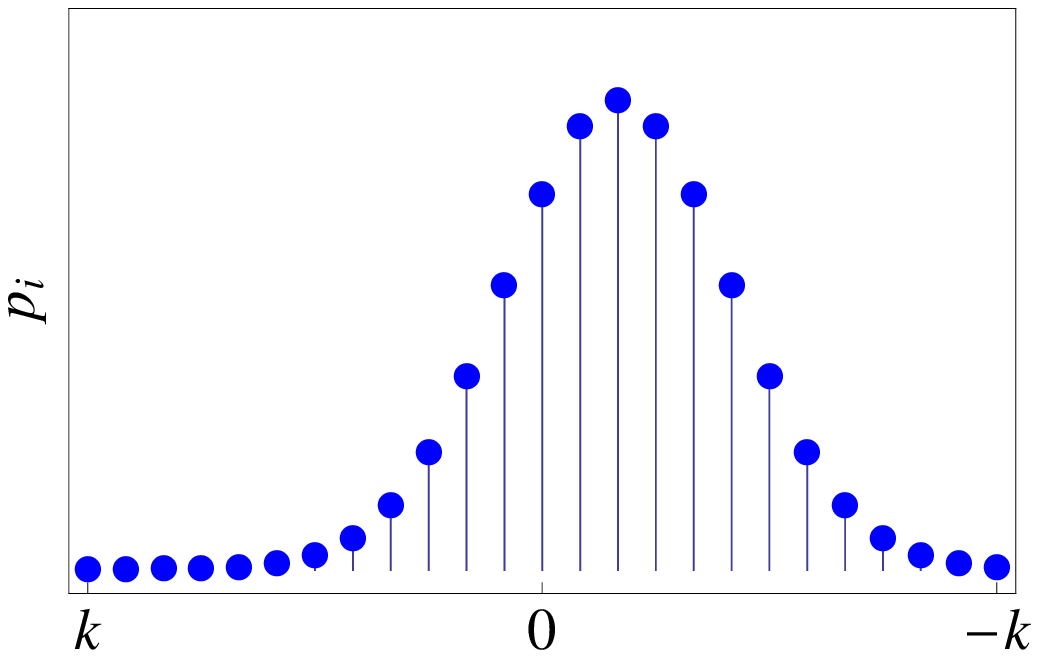}}
\end{center}
\caption{\label{gaussian} \it\small The gaussian model described in the text.}
\end{figure}
As $k$ goes to infinity we require
the probabilities $p_i$ approach the shifted gaussian distribution. In principle, all general results
about branching processes described below apply to this process as well. We will refer to this 
process as  the ``gaussian model", as opposed to the bacteria model discussed before. 
Clearly, the latter is much more tractable for explicit calculations.

Fortunately, for our purposes the two processes are equivalent. Indeed, let us start with a bacteria
process where at every time-step each bacterium divides into $N$ copies. Equivalently, one can consider another 
process, where at each step a bacterium divides into $\tilde{N}=N^k$ copies, which then hop to the $2k$
adjacent sites with probabilities $p_i$ generated by iterating $k$ times the elementary
process. Obviously, the late time properties of the two processes, such as the average number of dead
bacteria, or whether all bacteria die or not, are the same. The two processes are in fact one and the same---only in the latter case we are observing the system less frequently, every $k$ time-steps. Also, by the central limit theorem,
at large $k$ the probabilities $p_i$ generated in this way approach a gaussian distribution, so one gets the gaussian model.

Let us now consider the bacteria process and establish the exact relations between its parameters
$N$ and $p$ and the inflation parameters. As we said, it is enough to derive the 
diffusion equation and match its parameters in the two cases.
Also, from above we know that in the large $k$ limit the time interval $\Delta t$
corresponding to one step in the bacteria model goes to zero (while the time interval corresponding to
one step of the gaussian model $\widetilde{\Delta t} = k \Delta t$ remains constant). Let us consider one bacterium. The probability $P(j,n+1)$ for a bacterium to be at site $j$ at time step $(n+1)$ is given in terms of the probabilities at time $n$ by
%\begin{equation}
%\bar P(j,n+1)=\sum_i  \bar P(j,n+1|i,n) \bar P(i,n)\ ,
%\end{equation}
%where $\bar P(j,n+1|i,n)$ is the probability that the bacterium is at site $i$ at time step $n$. Using the fact that one bacterium can jump in one time step only to the two nearest sites, we obtain:
\begin{eqnarray}
\label{recursiveprob}
P(j,n+1)=(1-p)P(j-1,n)+p\;   P(j+1,n)\ .
\end{eqnarray}
We can now make the following identification
\begin{eqnarray}
j=-\frac{\phi}{\Delta\phi}\ , \quad \quad \quad n=\frac{t}{\Delta t} \;,
\end{eqnarray}
and treat $\phi$ and $t$ as continuous variables.
Taylor-expanding both sides of eq.~(\ref{recursiveprob}) up to linear order in $\Delta t$ and quadratic order in $\Delta \phi$
%write the correspondent equation for the function $\tilde P(\phi,t)\equiv P(-j(\phi),n(t))$ as:
%\begin{eqnarray}
%\tilde P (\phi,t+\Delta t)=(1-p)\tilde P(\phi+\Delta \phi,t)+p\; \tilde P(\phi-\Delta \phi,t)\ .
%\end{eqnarray}
%Expanding in $\Delta t$ and $\Delta \phi$, and dividing both members by $\Delta t$, 
we find
\begin{eqnarray}
\partial_t  P(\phi,t) \simeq (1-2p) \frac{\Delta \phi}{\Delta t} \, \partial_\phi  P(\phi,t) +\frac{1}{2}\frac{(\Delta \phi)^2}{\Delta t} \, \partial^2_\phi P(\phi,t)\;.
\end{eqnarray}
This equation can be expressed in terms of the variables $\psi=\phi-\dot\phi \, t$ and $\sigma^2=H^3 / 4\pi^2 \cdot t$ defined in sec.~\ref{sec:average}, obtaining
\begin{eqnarray}
\partial_{\sigma^2}  P(\psi,\sigma^2) \simeq \frac{4\pi^2}{H^3}\left((1-2p) \frac{\Delta \phi}{\Delta t}+\dot \phi\right)\partial_\psi \ P(\psi,\sigma^2) +\frac{1}{2}\frac{4\pi^2}{H^3}\frac{(\Delta \phi)^2}{\Delta t} \partial^2_\psi P(\psi,\sigma^2)\; .
\end{eqnarray}
This diffusion equation coincides with the inflationary one (\ref{diffusion}) if in the continuum
limit we take $\Delta\phi,\ \Delta t\to 0$ in such a way that
\be
\label{dtdf_relation}
\Delta t={4\pi^2\over H^3} (\Delta\phi)^2\;.
\ee
Also in this limit the hopping probability $p$ can be related to the slow
roll parameter $\Omega$ using $-(1-2p)\frac{\Delta \phi}{\Delta t}=\dot \phi$\;:
\be
\label{p_matching}
p={1\over 2}+{2\pi^2\dot{\phi}\over H^3} \, \Delta\phi={1\over 2}+\sqrt{6\pi^2\Omega} \, {\Delta\phi\over H}
\ee
%\begin{eqnarray}\label{eq: general matching}
%&&\frac{4\pi^2}{H^3}\frac{ \Delta \phi^2}{\Delta t}=1 \ , \\ \nonumber
%&&-(1-2p)\frac{\Delta \phi}{\Delta t}=\dot \phi\ .
%\end{eqnarray}
Finally, the number of bacteria copies $N$ at each reproduction event clearly is\begin{eqnarray}\label{eq: N matching}
N=1+3H \Delta t \;.
\end{eqnarray}
Given the discussion above one should not be surprised that the number of copies $N$ goes to
one in the continuum limit. This does not give rise to any problems, in spite of the fact that
strictly speaking the bacteria process is defined only for integer values of $N$. As we explained
the actual process we are interested in is the gaussian one with the number of copies
$\tilde{N}=N^k$ kept fixed in the continuum limit. One may regard working with the bacteria process and analytically continuing its results to real values of $N$ close to one as a technical trick to simplify the algebra for the gaussian process.

To directly confirm that the discrete approximation works let us start with reproducing with the bacteria model the results 
obtained in the previous sections. As a byproduct we will see at a more concrete level why the above analytic continuation works.

\subsection{\label{sec:bacteria_expectation_value} Dead-bacteria statistics: the average}
Let us start with the case of a semi-infinite line. Suppose we begin at $t=0$ with one single bacterium
at position $i$. Following ref.~\cite{branching_books}, for all different $i$'s we can define $M_{ij}$ as the average at the following time-step of the number of
bacteria at position $j$. If $i\neq 0$, the
bacterium produces on average $N p$ bacteria on its right and $N(1-p)$ on its left. If $i=0$, the bacterium
does not reproduce itself and stays there. Therefore, in matrix form $M_{ij}$ is given by
\begin{equation}\label{matrix M}
M= \left(
\begin{array}{ccccccc}
1 & 0 & 0 & 0 & \cdots\\
Np & 0 & N(1-p) & 0  & \cdots\\
0 & Np & 0 & N(1-p)  & \cdots\\
0 & 0  & Np & 0 &  \cdots \\
\vdots & \vdots & \vdots & \vdots &  \ddots
\end{array}
\right)\equiv\left(
\begin{array}{cc}
1 & 0 \\
I & \tilde M
\end{array}
\right) \ ,
\end{equation}
where we have represented $M$ in block-form defining
\be \label{eq:matrix tilde M}
I=\left(
\begin{array}{c}
Np \\
0 \\
0 \\
\vdots
\end{array}
\right) \ ,
\qquad
\tilde M=\left(
\begin{array}{cccccc}
 0 & N(1-p) & 0 &  \cdots\\
 Np & 0 & N(1-p) & \cdots\\
 0  & Np & 0 & \cdots \\
 \vdots & \vdots & \vdots & \ddots
\end{array}
\right)\ .
\ee
As the average is a linear operator, the matrix $M$ defines a linear map between the initial state vector and the average state after one time-step. That is, starting with $n_0, n_1, \dots$ bacteria at positions $0,1,\dots$, the average number of bacteria at position $j$ after  one time-step is $\sum_i n_i M_{ij}$.
We denote by $| i \ra$, for $i =0 ,1, \dots$ the elements of the canonical basis in which the above matrices have been written.

Likewise, consider the matrix $M^{(n)}_{ij}$ that gives after $n$ time-steps the average occupation number at position 
$j$ when starting with one bacterium at position $i$.
It is straightforward to show by induction that
$M^{(n)}= (M)^n$
\cite{branching_books}~\footnote{Usually in the branching-process
literature the ending site $i=0$ is not included, and  each bacterium has a given
probability of disappearing. This amounts to concentrating on the
matrix $\tilde M$. Here instead we include the occupation number of the
ending site so that we can follow the number of dead bacteria---the analogue of the reheating volume. It
is immediate to adapt the standard results to this case.}. Thanks to the block structure of $M$, we get
\be \label{eq: M^n}
M^{(n)}=\left(
\begin{array}{cc}
1 & 0 \\
\left(\sum_{m=0}^{n-1} \tilde M^m\right)|I\rangle & \tilde M^n
\end{array}
\right) \; .
\ee

We want to compute the average number of dead bacteria at
time $n$ starting with one bacterium at a generic site, and then
send $n$ to infinity. 
Suppose 
we start at the $i=1$ site (we will see below that this assumption does not affect our results). In this case we have to compute
\begin{equation}\label{eq:average}
\langle1|M^n|0\rangle=Np \, \langle1|\sum_{m=0}^{n-1}\tilde M^m
|1\rangle= Np \, \langle1|1+\tilde M+\tilde M^2+\ldots + \tilde
M^{n-1}|1\rangle \ .
\end{equation}

Restricting to the subspace spanned by the $| i \ra$'s with $i\ge 1$ we
define lowering and
raising operators $S$ and $S^\dag$ by
\begin{eqnarray}
S|i\rangle&=&\left\{ \begin{array}{ll} |i-1\rangle & \textrm{if
$i\neq 1$} \\
0 & \textrm{if $i=1$}
\end{array}
\right. \ , \\
S^\dag|i\rangle&=&|i+1\rangle \ .
\end{eqnarray}
It is easy to verify that these operators satisfy
\begin{equation} \label{Salgebra}
S\, S^\dag=1 \ , \qquad S^\dag S =1-|1\rangle\langle1|\ , \qquad
[S,S^\dag]=|1\rangle\langle1| \ .
\end{equation}
Then we can rewrite $\tilde M$ as 
\begin{equation} \label{tildeM def}
 \tilde M = N(1-p)\; S+Np\; S^\dag \ .
\end{equation}
Each term in eq.~(\ref{eq:average}) thus has the form
\begin{equation}
\langle1|\tilde M^m |1\rangle=\langle1|\left(N(1-p)S+Np S^\dag
\right)^m |1\rangle \ .
\end{equation}
This is non-zero only for even $m$, $m=2l$:
\begin{equation} \label{help1}
\langle1|\tilde M^{2l}|1\rangle=\left[N^2
p\;(1-p)\right]^{l} \, \langle1|\left(S+ S^\dag \right)^{2l}|1\rangle
 \equiv \left[N^2 p\;(1-p)\right]^{l} A^{(l)}
\; .
\end{equation}
Applying iteratively the commutation rule in eq.~(\ref{Salgebra}), we
can write
\begin{eqnarray}
A^{(l)}&=&\langle1|\left(S+S^\dag\right)\left(S+ S^\dag
\right)^{2l-1}|1\rangle= A^{(0)}A^{(l-1)}+\langle1|\left(S+S^\dag\right)S \left(S+
S^\dag \right)^{2l-2}|1\rangle  \nonumber
 \\ 
 &=& \! \! \ldots = A^{(0)}A^{(l-1)}+A^{(1)}A^{(l-2)}+\ldots+A^{(l-1)}A^{(0)}=\sum_{k=0}^{l-1}A^{(k)}A^{(l-1-k)} \; .
 \label{recursive1}
\end{eqnarray}
This rather complicated recursion relation can be solved by defining
the following generating function
\begin{equation}
F(x)=\sum_{i=0}^{\infty}A^{(l)}x^l\ , \qquad \textrm{so that} \;
\left.\frac{1}{l!}\frac{d^l F(x)}{dx^l}\right|_{x=0}=A^{(l)} \; .
\end{equation}
Then the above recursion relation translates into an algebraic, second order equation for $F$,
\begin{equation}
F(x)^2=\frac{F(x)}{x}-\frac{A^{(0)}}{x} \; ,
\end{equation}
whose
solution is
\begin{equation}
\label{Fx}
F(x)=\frac{1}{2x}\left(1\pm\sqrt{1-4x}\right) \; ,
\end{equation}
where we used $A^{(0)}=1$.
The solution relevant for us is that with the minus sign, since all the
$A^{(l)}$'s are positive. Expanding in Taylor series we find
\begin{equation}
A^{(l)}=2^l \frac{(2l-1)!!}{(l+1)!} \; .
\end{equation}

We are interested in studying the convergence of the average value of $V$, i.e.~the
number of dead bacteria at future infinity,
\be \label{Vaverage}
\la V \ra = Np \, \sum _{l=0} ^{\infty} \la 1 |  \tilde M^{2l} | 1 \ra \; ,
\ee
where we used eq.~(\ref{eq:average}).
At large $l$
\begin{equation}
A^{(l)}\sim  \frac{4^l}{l^{3/2}}  \; , \qquad \langle1|\tilde M^{2l}|1\rangle \sim \frac{\left[4 N^2p\;(1-p) \right]^{l}  }{ l^{3/2}}
\; .
\end{equation}
The series (\ref{Vaverage}) converges for
\begin{equation} \label{pc_average}
p \ge p_c \equiv \frac{1}{2}\left(1+\sqrt{1-\frac{1}{N^2}}\right) \ ,
\end{equation}
and it diverges for smaller $p$'s. By making use of the matching relations (\ref{dtdf_relation}--\ref{eq: N matching}) one immediately checks that in the continuum 
limit the expression for $p_c$ agrees with the correct inflationary result for $\Omega$, $\Omega=1$. 

This result does not depend on the initial position of the original bacterium, which in the above we assumed to be $i=1$. In fact, any other site $j$ can be reached starting from $i$ in $(j-i)$ time-steps with finite probability, and vice-versa. This implies that if the average of $V$ diverges starting from a given position, it must also diverge starting from any other position. Of course this conclusion also holds for all higher moments of $V$, so in the following sections we will always assume that we start from $i=1$.

Note finally, that (\ref{Vaverage}) implies that at $p>p_c$, when $\langle V\rangle$ is finite, all
eigenvalues of $\tilde{M}$ are smaller than one, and at $p<p_c$ at least one eigenvalue is larger than
one. Indeed, a well-known result in linear algebra is that an arbitrary matrix $\tilde{M}$ can be presented in the form
\[
\tilde M=A ^{-1}JA\;,
\]
where $A$ is a non-degenerate matrix and $J$ has the normal Jordan form ({\em i.e.}, its only non-zero elements are those on the main diagonal---equal to the eigenvalues $\lambda_i$ of the matrix $\tilde M$---and those on the diagonal right above the main one---equal to one). Consequently, calculating powers of $\tilde{M}$
reduces to taking powers of $J$, that have a rather simple form, and all matrix elements of $J^{2l}$
scale as $\lambda_{i}^{2l}$ at large $l$.
Consequently, the sum (\ref{Vaverage}) diverges iff one of the eigenvalues $|\lambda_i|>1$~\footnote{This is assuming that the vector $|1\rangle$ has a non-trivial projection onto the corresponding Jordan block. This is true in our case, as follows from the above argument that if (\ref{Vaverage}) diverges for 
some initial position $i$, then it diverges for all other initial positions as well.}.

Notice now, that considering an extended bacteria process with a non integer number of children $N$ instead of the gaussian one corresponds to working
with the matrix $\tilde M$ with non integer $N$ instead of $\tilde M^k$ with $k$ such that $N^k=\tilde N$ is constant and integer. Clearly, this doesn't change whether there is 
an eigenvalue larger than one, so the critical probabilities $p_c$ are
the same in the two cases.
%~\footnote{In fact, one can look at the fact that we are able to reproduce the critical value of the divergency of $\langle V \rangle$ as a verification that there are no hidden discontinuities in taking the limit $k\rightarrow \infty$, $\Delta t\rightarrow 0$ from the extended bacteria process to the gaussian process.}.

\subsection{\label{sec:bacteria_variance_expectation_value}The variance: enhancement for
the infinite line}

We now consider $\la V^2 \ra$. To compute it, it is convenient to introduce the matrix of second moments for the bacteria populations at different sites. Namely, assuming for simplicity that we start with one bacterium at the $i=1$ site, we define $C^{(n)} _{ij} \equiv \la N_i^{(n)}  N_j^{(n)} \ra$, where $N^{(n)}_i$ is the number of bacteria at site $i$ after $n$ time-steps.
Then we are interested in
\be \label{v2}
\la V^2 \ra = \lim_{n \to \infty} C^{(n)}_{00} \; . 
\ee

An explicit expression for $C^{(n)}$ can be obtained recursively \cite{branching_books}. Using
the  generating function technique reviewed in the next section
it is straightforward to show that
\be \label{discrete_variance}
C^{(n + 1)}= M^T C^{(n)}  M + \sum_{i=0}^\infty W_{(i)} \langle 1|M^n | i \rangle   \; ,
\ee
where $M$ is the matrix of averages (\ref{matrix M}), and $W_{(i)}$ is the covariance matrix of bacteria after one time step if one starts with one bacterium at site $i$. The sum runs over all sites; once again we assume that we have a semi-infinite line of sites.
The above recursion relation can be easily solved to yield \cite{branching_books}
\begin{equation} \label{Cn}
C^{(n)}= \left(M^T\right)^n C^{(0)}  M^n + \sum_{m=1}^n \left(M^T\right)^{n-m}
\l\sum_{i=0}^\infty W_{(i)} \langle 1|M^{m-1} | i \rangle 
%\left(M^{m-1} \right)_{1 i}
\r
M^{n-m} \; .
\end{equation}

We want to show that the variance (\ref{v2}) as a function of $p$  diverges at a critical value $p^{\rm var}_c$ where the average is still finite, that is $p^{\rm var}_c > p_c$. For this purpose a lower bound on eq.~(\ref{v2}) suffices. Since all terms in eq.~(\ref{Cn}) are positive definite, we can neglect the first term and consider just the sum over $m$. 

The matrix $W_{(i)}$ is straightforward to compute. Starting from the $i$-th site with one bacterium, 
the numbers of bacteria at nearby sites after one time step obey a binomial distribution. Obviously $W_{(0)} = 0$, because bacteria at $i=0$ neither reproduce themselves nor move around,  whereas for $i \ge 1$ we get
\be \label{V_i_help}
W_{(i \neq 0)}= Np\;(1-p)\begin{array}{ccccc} \left(
\begin{array}{ccccccc} 
 \ddots 	&	&	& \stackrel{\displaystyle{i}}{\downarrow}	&	&	&	\\
 		& 0	& 0 	& 0 	& 0 	& 0	& 	\\
 		& 0	& 1 	& 0 	& -1 	& 0	&   	\\
 i \to		& 0	& 0 	& 0 	& 0 	& 0	&  	\\
 		& 0	& -1 	& 0 	& 1 	& 0	&   	\\
		& 0	& 0 	& 0 	& 0 	& 0 	& 	\\
		& 	&	&	&	& 	& \ddots
\end{array}
\right)
\end{array} \equiv 2 N p\;(1-p)\;|\psi_i\rangle\langle \psi_i | \ ,
\ee
that is, $W_{(i)}$ is proportional to the projector on the state 
\begin{equation}
| \psi_i \rangle  \equiv  \frac{1}{\sqrt{2}}\big(|i-1\rangle-|i+1\rangle\big) \ .
\end{equation}
We then have
\begin{eqnarray}
C_{00}^{(n)} & \ge &  \langle0| \sum_{m=1}^n \left(M^T\right)^{n-m}
\bigg(\sum_{i=0}^\infty W_{(i)} \langle 1|M^{m-1} | i \rangle
% \left(M^{m-1} \right)_{1 i}
 \bigg) M^{n-m} |0\rangle  = \nonumber \\
\label{eq:evident_variance}
& = & 2 N p \, (1-p) \,
\sum_{m=1}^n \sum_{i=0}^\infty
\langle 1|M^{m-1} | i \rangle \ 
 %\left(M^{m-1}\right)_{1i}  
 |\langle\psi_i | M^{n-m} |0\rangle|^2 \; .
 \label{C00}
\end{eqnarray}

All terms in the sum are positive, so that for getting a lower bound we can restrict to any of them. 
For $n$ even we just consider the term with $m = i = n/2$. Using the lowering and raising operators introduced in the previous section it is straightforward to show that
\begin{eqnarray}  
&& \langle \psi_{n/2}|M^{\frac{n}{2}}|0\rangle=\sfrac{1}{\sqrt{2}}\left(Np\right)^{\frac{n}{2}-2}
\langle n/2-1|\left(S^\dag\right)^{\frac{n}{2}-2}|1 \rangle
=\sfrac{1}{\sqrt{2}}\left(Np\right)^{\frac{n}{2}-2}
 \ , \\
&& \langle 1|M^{{n\over 2}-1} | {n/2}\rangle 
%\left(M ^{\frac{n}{2}-1} \right)_{1 \frac{n}{2}}
= 
\big(N(1- p)\big)^{\frac{n}{2}-1}
\langle 1| \, S^{\frac{n}{2}-1} \, | n/2 \rangle= \big(N(1-p)\big)^{\frac{n}{2}-1} \ .
\end{eqnarray}
Plugging these results back into eq.~(\ref{eq:evident_variance}) we get
\begin{equation}
C^{(n)}_{00} \ge \frac{1}{N^4 p^3} \left(N^3 p^2 (1-p)\right)^{n/2} \ .
\end{equation}
As $n\rightarrow \infty$, this quantity diverges for
\begin{equation}
\label{easy_bound}
N^3 p^2 (1-p)>1 \; , \qquad \textrm{ or } \ p<p_c^{\rm cov} \  \textrm{
with }\  p^{\rm cov}_c\gtrsim1-\frac{1}{N^3} \ \textrm{ for } N\gg
1 \ .
\end{equation}
As claimed, we notice that $p_c^{\rm cov}>p_c$ for the infinite
chain: the covariance of the number of dead bacteria diverges
before the expectation value. This qualitatively agrees with the inflationary case, however
at the quantitative level the bound (\ref{easy_bound}) is too weak---for instance, it is useless in the continuum limit,
$N\to 1$.

To exactly reproduce the inflationary results note that the sum over the site number
$i$ in eq.~(\ref{discrete_variance})  is the discrete analogue of the convolution over the inflaton field values at horizon crossing in eq.~(\ref{convolution}). Similarly, the discrete variable $m$ in
(\ref{discrete_variance}) corresponds to the horizon crossing time $t_*$ in the continuum case.
Both integrals, over $t_*$ and over inflaton values at horizon-crossing, were dominated by the saddle point, determined by eq.~(\ref{t*min}) and eq.~(\ref{eq:saddleback}). This suggests that the double 
sum in eq.~(\ref{discrete_variance}) is also dominated by a single term. The natural guess  for the optimal choice\footnote{Using the 
approximate expressions we derive below it is straightforward to check directly that this is indeed the optimal choice.} for $m$ based on
 eq.~(\ref{t*min})  is 
\be
\label{bestm}
m_*=n\l\sqrt{2\Omega}-1\r\;,
\ee
where we identified the total number of steps with the total time, so that $n=t_+/(2\Delta t)$. Similarly, from
eq.~(\ref{eq:saddleback}) we deduce  the optimal choice for $i$,
\be
\label{besti}
i_*=\l \sqrt{2/\Omega}-1\r \dot{\phi} \, m_*{\Delta t\over\Delta\phi}=2\pi\sqrt{3}\l 3\sqrt{2\Omega}-2-2\Omega \r
H^{-1}n\Delta\phi\;,
\ee
where at the last step we used the matching relations (\ref{dtdf_relation}--\ref{eq: N matching}). 
To calculate the corresponding term in eq.~(\ref{discrete_variance})
we need expressions for the matrix elements $\langle 1|M^{m_*-1}|i_*\rangle$ and
$\langle\psi_{i_*}|M^{n-m_*}|0\rangle$. Actually, as before, to decide whether the variance
remains finite in the late time (large $n$) limit we need just the leading exponential
asymptotics of these elements. In this limit one can replace these
elements by  $\langle 1|\tilde{M}^{m_*}|i_*\rangle$ and
$\langle i_*|\tilde{M}^{n-m_*}|1\rangle$. Then  by making use of eq.~(\ref{tildeM def}) we write
\be
\label{Mmi1}
\langle 1|M^{m_*-1}|i_*\rangle\simeq\langle 1|\tilde{M}^{m_*}|i_*\rangle
%\simeq\langle 1|(N(1-p)S+NpS^\dag)^{m_*}|i_*\rangle
=
N^{m_*}(1-p)^{m_*+i_*-2\over 2}p^{m_*-i_*-2\over2}\langle 1|(S+S^\dag)^{m_*}|i_*\rangle
\ee
for even $(m_*+i_*)$, and zero otherwise.
In the leading exponential approximation one can estimate $\langle 1|(S+S^\dag)^{m_*}|i_*\rangle$
just by counting the monomials in the expansion of $(S+S^\dag)^{m_*}$ that contain $(i_*-1)$ more $S$'s than $S^\dag$'s. This gives
\be
\label{binom_estimate}
\langle 1|(S+S^\dag)^{m_*}|i_*\rangle\simeq
\binom{m_*}{{m_*+i*-2\over 2}}
\simeq
\l2^\mu\mu^\mu(\mu^2-1)^{-\mu/2}
\sqrt{\mu-1\over\mu+1}
\r^{i_*}\;,
\ee
where $\mu=m_*/i_*$ does not depend on $n$ and is determined by (\ref{bestm}) and (\ref{besti}).
The validity of the estimate (\ref{binom_estimate}) is not immediately obvious, because many of the
monomials with the right number of $S$ and $S^\dag$ still give vanishing matrix elements.
Using techniques similar to those used in section~\ref{sec:bacteria_expectation_value}
one can check that this estimate is nevertheless correct (see Appendix~\ref{sheets}). The same
logic applied to the matrix element $\langle\psi_{i_*}|M^{n-m_*}|0\rangle$ gives
\be
\langle\psi_{i_*}|M^{n-m_*}|0\rangle
%\simeq\langle i_*|\tilde{M}^{n-m_*}|1\rangle
\simeq
N^{n-m_*}p^{n-m_*+i_*-2\over 2}(1-p)^{n-m_*-i_*-2\over2}
\binom{n-m_*}{{n-m_*+i*-2\over 2}}
\ee
Finally, by making use of these equations and of the expressions (\ref{bestm}), (\ref{besti}) for $m_*$, $i_*$
one obtains
\be
\label{awesomeO}
\langle 1|M^{m_*-1}|i_*\rangle|\langle\psi_{i_*}|M^{n-m_*}|0\rangle|^2\simeq
\l1-12\pi^2\l{\Delta\phi\over H}\r^2{4\sqrt{2}\Omega^{3/2}-18\Omega+13\sqrt{2\Omega}-6\over
2\Omega-3\sqrt{2\Omega}+2}+{\cal O}\l\Delta\phi^3\r\r^n
\ee
This contribution (and, consequently, the variance of $V$) diverges at late times
when the coefficient of the $(\Delta\phi)^2$ term crosses zero. This happens at $\Omega=9/8$, so we exactly reproduce the inflationary result for when the variance diverges, eq.~(\ref{epscr}). 
Note that, just like in the continuous case, this result crucially relies on the infinite range for the
inflaton field (infinite number of sites in the bacteria model). Indeed, for a finite chain
the sum over $i$ in (\ref{C00}) runs up to some finite value $i_{max}$, so that at large enough $n$
one has  $i_*>i_{max}$ analogously to how in the continuous case the saddle point (\ref{eq:saddleback}) ceases to be in the integration region. Moreover, analogously to the case of the average, it is clear from eqs.~(\ref{Cn}) and 
(\ref{C00})  that the variance remains finite if  all eigenvalues of the matrix of averages $\tilde{M}$ are
smaller than unity for the finite chain, so that the variance and the average diverge at the same value of
$p$. This is true for all higher  moments as well (see, e.g. \cite{branching_books})

\subsection{Extinction probability in the semi-infinite line
\label{sec:baceteria_normalization}}

The principal advantage of the bacteria model is that it allows to directly study the transition
to the eternal  regime, without relying on indirect criteria, such as the divergence of  the probability distribution moments for the number of dead bacteria (reheating volume in the
inflationary case). Indeed the most direct and physical characterization of eternal inflation
is that it can run {\it forever}. In the context of the bacteria model this corresponds to a non-zero
probability that the population never dies out. 

There is a theorem (see e.g.~\cite{branching_books}) that for a finite chain the extinction probability is equal to one for a jumping probability
$p$ close to one. However, when $p$ drops below a certain critical value $p_c$ the extinction probability becomes smaller than one. At the critical value $p_c$ one of the eigenvalues of the
matrix of averages $\tilde M$ crosses one and stays larger than one in the eternal regime.
In the previous two sections we saw that when the  matrix $\tilde M$ develops
an eigenvalue larger than 1, then the expectation value and the
variance of the number of dead bacteria begin to diverge as time
$n$ goes to infinity. We see now that at the {\it same}
critical value of $p$ the extinction probability drops below one. Given the importance of this result,
we find it instructive to explain it in more detail; as a byproduct we will see 
that the whole picture can be continuously extended to the infinite line case. This will confirm our expectation that also in the infinite line case, the transition
to the eternal regime happens at the same value of $p$ (slow roll parameter in the inflationary case) where the average number of dead bacteria diverges.

Let us first consider a branching process on a line of length $L$. 
 A convenient tool to study the branching process is the set of generating functions $f^{(n)}_i (s_j)$, where
 $i,j=1,\dots, L$. These are defined as power series
 \be
 \label{fs_def}
 f^{(n)}_i(s_j)=\sum_{k_1\dots k_L} p^{(n)}_{i; k_1\dots k_L}s_1^{k_1}\dots s_L^{k_L}
 \ee
where $p^{(n)}_{i; k_1\dots k_L}$ is the probability that in a branching process that started with a single
bacterium at the $i$-th site after $n$ steps one has $k_1$ bacteria at the first site, $k_2$ bacteria
at the second site, etc.
 It is convenient to combine together 
 all functions $f^{(n)}_i$ with the same number
of steps $n$ into a map $F_n$ from the $L$-dimensional space of the auxiliary parameters $s_i$
into an $L$-dimensional space parametrized by the $f_i$'s. Also in what follows we often drop
the subscript from the $s_i$ variables and denote by $s$ a point in the $L$-dimensional space
with coordinates $(s_1,\dots, s_L)$.

The main property making generating functions useful is the iterative relation
\be
\label{iterative}
F_{n+1}=F_1(F_n)\;.
\ee
This property is straightforward to check by making use of the definition of the
branching process and elementary 
properties of probabilities.
We are interested in the late time behavior of the branching process, which is determined
by the limiting function $F_{\infty}$. The iterative property (\ref{iterative}) implies that 
\[
F_1(F_\infty)=F_\infty\;,
\]
{\it i.e.}~the set of values of the function $F_\infty$ is a subset of the fixed points
of the function $F_1$, such that
\be
\label{fixed}
F_1(s)=s\;.
\ee
For our purposes is enough to study the mapping $F_1$ inside the $L$-dimensional cube $I_L$ of unit size,
$0\leq s_i<1$. The definition (\ref{fs_def}) of the generating functions implies that all partial derivatives of
$F_1(s)$ are positive. Also, the normalization of probabilities implies that
\[
F_1(1,\dots,1)=(1,\dots, 1)\equiv \vec{1}\;.
\]
Another important property of the mapping $F_1$, also immediately following from its definition,
 is that the Jacobian of $F_1$ at the 
point $s=(1,\dots,1)$ is equal to the matrix of averages $\tilde{M}$. Clearly, the mappings $F_n$ also
satisfy the straightforward analogues of all these properties.
This and the iterative relation (\ref{iterative}) immediately implies the relation
\[
\tilde{M}^{(n)}=\tilde{M}^n
\]
used above to calculate the late time behavior of the average number of the dead bacteria.
Also by applying the chain rule twice to (\ref{iterative}) it is straightforward to reproduce the recursion relation  (\ref{discrete_variance}) used to calculate the variance.

We are ready now to discuss how the transition to the eternal regime happens. Note first, that if 
the mapping $F_1$ has no other fixed points in the cube $I_L$ apart from $\vec{1}$ (see fig.~\ref{extinction}), then 
\[
F_\infty=\vec{1}\;.
\]
By definition of the generating functions, eq.~({\ref{fs_def}}), this means that in the late time asymptotics with
probability one there are no bacteria at any of the sites. The extinction probability is exactly equal to
one (inflation ends). The situation changes when  a non-trivial fixed point $s_f$ solving eq.~(\ref{fixed}) enters the region $I_L$ (see fig.~\ref{extinction}). 
\begin{figure}[b!]
\begin{center}
\includegraphics[width=13cm]{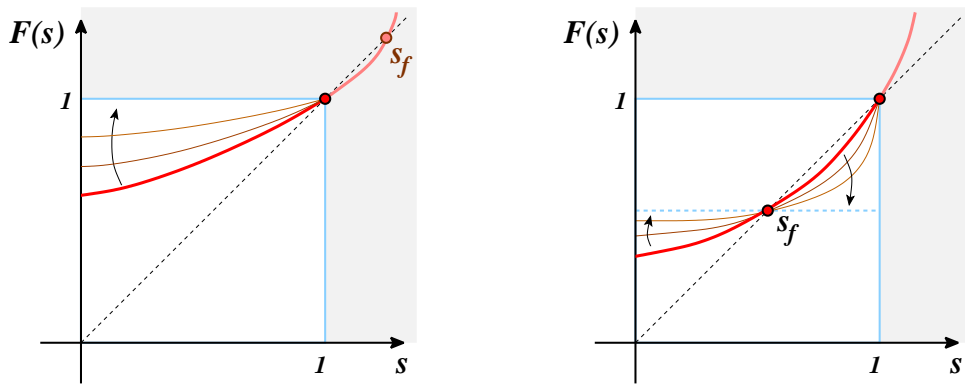}
\caption{\label{extinction} \small \it {\em Left:} Plot of $F_1(s)$ for large $p$ {\em (thick curve)}. The only fixed point  in the unit cube is $s=1$. Further applications of $F_1$ {\em (thinner curves)} drive the curve to the $F_{\infty} = 1$ line. {\em Right}: For smaller $p$'s a new fixed point $s_f$ enters the unit cube.  Now the limiting line is $F_{\infty} = s_f$.}
\end{center}
\end{figure}
Now one has 
\[
F_\infty=s_f<\vec{1}\;.
\]
This implies that, as before, the probability to have any finite non-zero  number of bacteria at any site vanishes. However, the probabilities to have zero bacteria at the various sites, 
\be
p^{(\infty)}_{i;0\dots0} = f_i^{(\infty)}(0)=(s_f)_i \; ,
\ee
are all less than one.  This means that there is a non-vanishing probability that the population never dies out  and that the total number of bacteria grows indefinitely
at late times. This corresponds to the eternal inflation regime. Clearly, this implies that
the number of dead bacteria also has a finite probability to grow indefinitely; in the context of eternal inflation this translates into
\[
1-\int\rho(V)dV>0\;.
\]
%To see it directly one may repeat the above arguments with an extra site corresponding to the dead bacteria
%added, as we did in  section~\ref {sec:bacteria_expectation_value} (see also an explicit example
%in section~\ref{sec:time_dead_probability}).

To identify the critical value of $p$ where the transition happens it is instructive to track the position of the fixed point $s_f$ in $s$-space
as a function of  $p$. According to the above discussion,
at the critical
value of $p$  the fixed point enters into the unit cube $I_L$. It is straightforward to check that the
monotonic property of the generating functions (\ref{fs_def}) imply that  the trajectory
of the fixed point
necessarily passes through the point $\vec{1}$. Consequently, at the critical probability the graph of the
mapping $F_1(s)$ is tangent to the surface $f_i=s_i$ at $\vec{1}$ along one direction in the $(2L)$-dimensional space
parametrized by the $f_i$'s and $s_i$'s. The projection of this direction onto the $s$-hyperplane is tangent
to the trajectory of the fixed point at $\vec{1}$. Algebraically this means that at the
transition $p$ one of the eigenvalues of the Jacobian of $F_1$ at $\vec{1}$ is equal to one,
\be
\label{Jacobian_eigen}
{\d F_1(\vec{1})\over \d s}v\equiv
{\d f_i^{(1)}(\vec{1})\over \d s_j}v_j=v_i\;,
\ee
where $v$ is a vector tangent to the fixed-point trajectory at $\vec{1}$.
At lower values of $p$ this eigenvalue becomes larger than one.
Recalling that the Jacobian of $F_1$ at $\vec{1}$ coincides with the matrix of averages $\tilde{M}$, we see that the
transition to the eternal regime indeed happens at the same value $p_c$ where the average number of dead bacteria (and, in the finite line case, all higher moments) diverges at late times.

Let us now extend this result to the infinite line.
The infinite-line process can be approximated by a sequence of finite-dimensional branching processes with defining matrices $\tilde M_{L}$, where $\tilde M_L$ is the projection of $\tilde M$ onto the first $L$ basis vectors. The matrix $\tilde M_L$ describes a branching process with $L$ sites, where bacteria can also ``die''  with probability $(1-p)$ jumping from the $L$-th site to the left. 
%This is slightly different from the finite-dimensional branching process we studied in the last section, but all the generic theorems  still hold.
Now, the crucial point is that the extinction probability for the infinite line is smaller than that of the branching process described by $\tilde M_L$, for any finite $L$. This is quite clear as in going from the finite to the infinite case we are increasing the survival probability of each bacterium. Therefore if we are able to prove that for $p<p_c$ and sufficiently large $L$ the extinction probability associated with $\tilde M_L$ is strictly less than one, our claim follows. The finite-dimensional theorem implies that this is equivalent to showing that for sufficiently large $L$ the maximum eigenvalue of $\tilde M_L$ is larger than one.

The matrix $\tilde M_L$ is just the $L\times L$ truncation of $\tilde M$ in eq.~(\ref{eq:matrix tilde M}),
\be
\tilde M_L=\left(
\begin{array}{ccccc}
 0 & N(1-p) & 0 &  \cdots & \\
 \; \; Np \; \; & 0 & \ddots &  & \vdots\\
 0  & \ddots & \ddots &  & 0 \\
 \vdots & &  & 0 & N(1-p) \\
 & \cdots & 0 & \; \; Np \; \; & 0
\end{array}
\right)_{L\times L}\ .
\ee
We want to solve the eigenvalue problem $\la v | \tilde M_L  = \lambda \la v |$. In components this reads
\begin{eqnarray}
&&N p \, v_2=\lambda \, v_1 \label{eq:rec 1}\\
&&Np\, v_{j+1}+N (1-p) \, v_{j-1}=\lambda\, v_j \qquad  \textrm{ for } \ 1 < j < n \label{eq:rec 2} \\
&& N (1-p) \, v_{L-1} = \lambda\, v_L \ . \label{eq:rec 3} 
\end{eqnarray}
Let's try the  ansatz
\begin{equation}
v_j=A^j \; ,
\end{equation}
where $A$ is a complex number. Eq.~(\ref{eq:rec 2}) implies
\be \label{Apm}
N p\, A^2 -\lambda\, A +N(1-p)=0 \qquad \Rightarrow \qquad A_\pm = \frac{\lambda\pm\sqrt{\lambda^2-4N^2p(1-p)}}{2Np} \; . 
\ee
Eq.~(\ref{eq:rec 1}) is equivalent to the recursion relation eq.~(\ref{eq:rec 2}) with initial condition $v_0 =0$. This then implies that the correct linear combination is
\be
v_j = A_+^j - A_-^j \; .
\ee
Analogously, eq.~(\ref{eq:rec 3}) gives the `final condition' $v_{L+1} = 0$,
\be \label{phase}
A_+^{L+1} = A_-^{L+1} \;. 
\ee
This cannot hold if $A_{\pm}$ are distinct real roots. Then the square root in eq.~(\ref{Apm}) must be purely imaginary, $\lambda^2 \le 4N^2p(1-p)$, in which case $A_+$ and $A_-$ are complex-conjugate of each other. Therefore eq.~(\ref{phase}) implies that
\be \label{kp}
{\rm arg}(A_+^{L+1}) = k \cdot \pi  \qquad \Rightarrow \qquad  {\rm arg}(A_+) = k \cdot \frac{\pi}{L+1} \; ,
\ee
where $k$ is an integer. Distinct solutions correspond to $1 \le k \le L$, which exhaust the set of eigenvalues of $\tilde M_L$.
From eq.~(\ref{Apm}) we have
\be
\cos\big( {\rm arg}(A_+) \big)= \frac{\lambda}{\sqrt{4N^2p(1-p)}} \; ,
\ee
which combined with eq.~(\ref{kp}) finally yields the eigenvalues
\be
\lambda_k = \sqrt{4N^2p(1-p)} \cdot \cos \bigg( \frac{\pi \, k}{L+1} \bigg) \; , \qquad 1 \le k \le L \; .
\ee
The largest eigenvalue is that with $k=1$, which for sufficiently large $L$ is arbitrarily close to $\sqrt{4N^2p(1-p)}$. For $p < p_c$, $\sqrt{4N^2p(1-p)} > 1$ (see eq.~(\ref{pc_average})), and the largest eigenvalue will become strictly larger than one for sufficiently large $L$. This completes the proof. 

\subsection{\label{sec:time_dead_probability} How the probability distribution can lose its normalization}

To understand how the extinction probability can change abruptly when $p$ crosses $p_c$,
it is instructive to track the probability distribution for the number of dead bacteria as a function of time.
For simplicity we do this for a minimal branching process with just two sites and $N=2$ copies at each reproduction event (see fig.~\ref{fig:2_site_line}).
\begin{figure}[b!]
\begin{center}
\includegraphics[width=3.4cm]{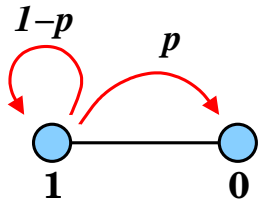}
\caption{\label{fig:2_site_line} \it \small The 2-site branching process.}
\end{center}
\end{figure}
In this case the generating functions (\ref{fs_def}) are particularly simple:
\begin{eqnarray}
&&f^{(1)}_0 (s_0,s_1)=s_0 \ , \\
&&f^{(1)}_1 (s_0,s_1)=\big((1-p)s_1+p s_0 \big)^2 \ .
\end{eqnarray}
It is straightforward to apply here the generic results discussed in the last section. The critical probability is the value of $p$ for which $\partial _{s_1} f^{(1)} _1$ computed at $(s_0, s_1) = (1,1)$ becomes larger than one. We get
\be
p_c = \frac12 \; .
\ee
The extinction probability is the fixed point of $f^{(1)} _1$ (setting $s_0 = 1$) inside the unit interval $0 \le s_1 \le 1$,
\be
p_{\rm ext} = \frac{ 1-2p + 2p^2 - \sqrt{(1-2p)^2} }{ 2(1-p)^2 } =
\left\{ \begin{array}{ccc}
	1 & & p > 1/2 \\
	(\frac{p}{1-p})^2 & & p < 1/2
	\end{array} \right. \; ,
\ee
which indeed drops below one for $p<p_c$.

To follow the time-evolution of the dead-bacteria probability distribution we numerically iterate the recursion relation $F_{n+1}=F_1(F_n)$ for different values of $p$. As before $n$ counts the number of time-steps.
The results are plotted in fig.~\ref{fig:subcritical}. 
\begin{figure}[t!]
\includegraphics[width=8.4cm]{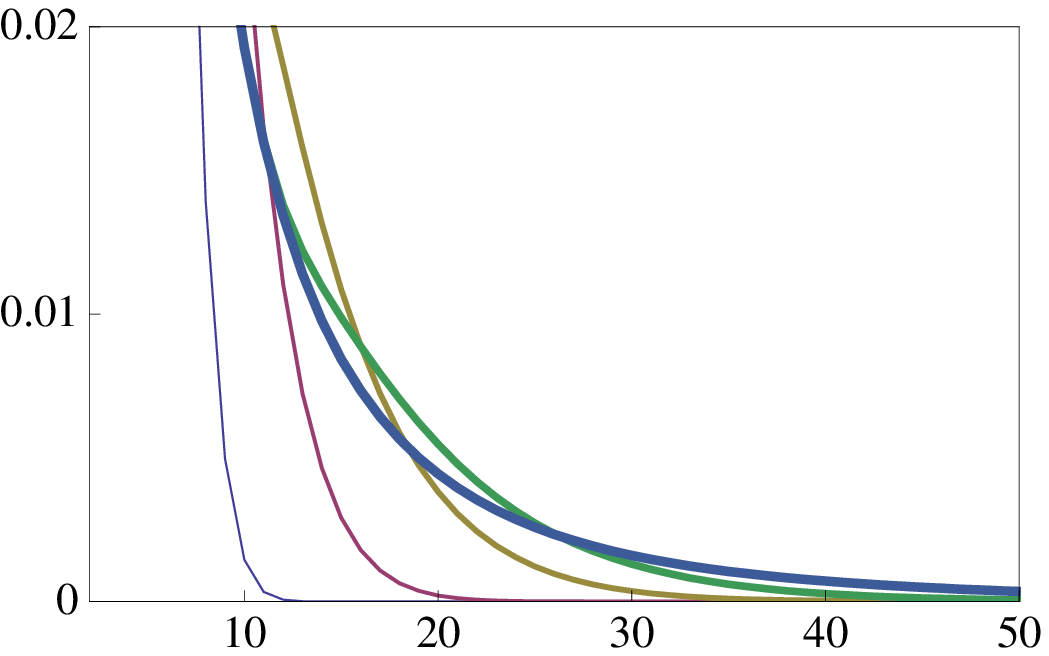}
\hfill
\includegraphics[width=8.4cm]{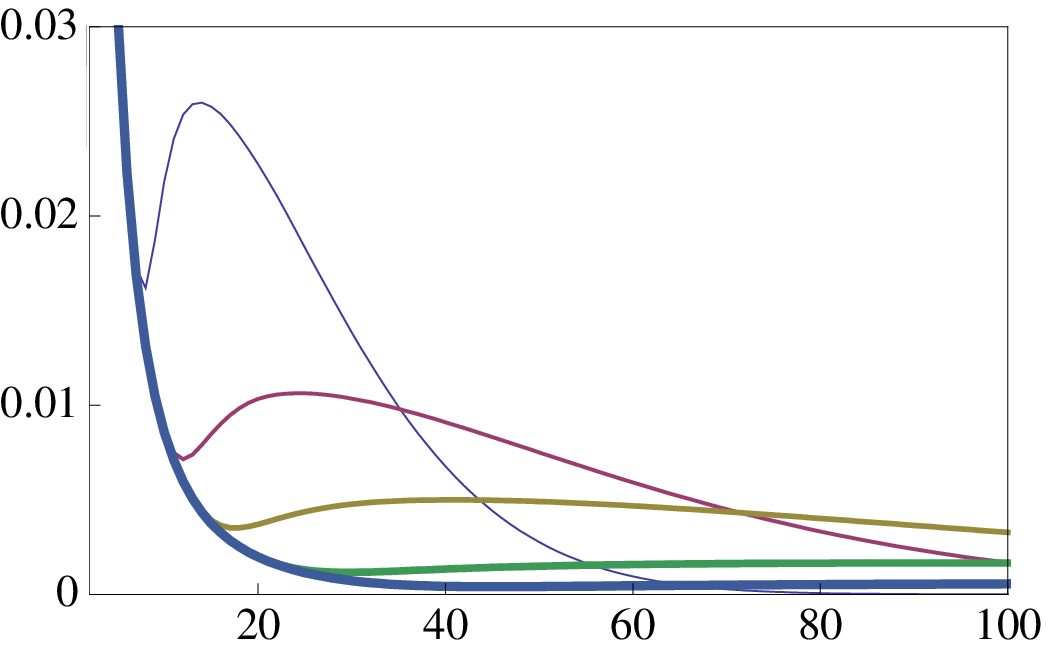}
\caption{\label{fig:subcritical} \it \small Time evolution of the dead-bacteria probability distribution in the two-site model. Thinner (thicker) lines correspond to earlier (later) times. {\em Left:} supercritical case, $p > p_c$. As a function of time the distribution boringly relaxes to its asymptotic form. {\em Right:} subcritical case, $p < p_c$. A ``bump'' develops and gradually shifts towards larger and larger dead bacteria counts, while getting wider and lower, and bringing some finite probability to infinity.}
\end{figure}
For  $p$ larger than the critical probability, the probability distribution at early times is strongly peaked around small numbers, and as time goes on it smoothly asymptotes to its 
$n \to \infty$ form. The probability distribution is obviously normalized to one for every $n$, and so is the asymptotic $n \to \infty$ distribution. Related to this, the extinction probability is one, as expected.

More interesting is the $p < p_c$ case. There, at early times one can notice a feature---a ``bump''---in the probability distribution. As time goes on the bump moves towards larger and larger dead-bacteria numbers, spreading out at the same time. The $n \to \infty$ asymptotic distribution shows no bump. Still, one can check that its normalization is strictly smaller than one---it is in fact the extinction probability $p_{\rm ext} < 1$, as it should.
The bump has taken away some finite probability to infinity! The fact that the asymptotic probability distribution is not normalized to one is due to the fact there is a finite probability of having infinitely many dead bacteria.

We warn the reader that the qualitative presence of the bump is not a distinctive feature of the subcritical ($p<p_c$) case. The plots in fig.~\ref{fig:subcritical} were done with $p=0.6$ and $p=0.4$. However a bump can also be noticed in the supercritical case, $p>p_c$, if we take $p$ quite close to $p_c$, say $p=0.51$. In such a case the bump follows the same qualitative time-evolution as for the subcritical case---it moves to the right, spreads out, and disappears. The question of whether it takes away some finite probability to infinity is then a quantitative one, and an explicit check shows that is does not, as expected.

Technically what happens in the $p < p_c$ case is that as $n \to \infty$ the sequence of probability distributions $P_n(V)$ converges {\em pointwise} to the asymptotic $P_{\infty} (V)$, 
\be
\label{point}
\forall V \qquad P_n (V) - P_{\infty} (V) \to 0 \; ,
\ee
but it does not converge {\em strongly},
\be \label{strong}
|| P_n (V) - P_{\infty} (V) || \to (1 - p_{\rm ext}) \neq 0 \; ,
\ee
where the relevant norm here is the $L^1$ norm,
$|| f ||_1 \equiv \int \! dx \: |f(x)| $~\footnote{We are using a continuous notation for $V$, because we have in mind the reheating volume in inflation, whereas in fact in the model at hand $V$ is a discrete variable---the number of dead bacteria. The proper changes in the notation are obvious.}.
As a consequence taking the $n \to \infty$ limit and taking the norm do not commute. The r.h.s~of eq.~(\ref{strong}) is precisely the probability taken away by the ``bump''.

However we can formally supplement the asymptotic probability with a $\delta$-function supported at infinity, with coefficent $(1-p_{\rm ext})$,
\be \label{deltafunction}
P_{\infty} (V) = \lim_{n\to\infty} P_n (V) + (1-p_{\rm ext}) \, \delta(V -\infty) \; .
\ee
Such a term fixes the normalization, and it also makes all moments diverge, as we know they must. 
Indeed if one does not include the point at infinity, the moments
computed with the asymptotic probability (\ref{point}) do not coincide with those we computed in the previous sections,  which read
\begin{equation}
\langle V^m \rangle= \lim_{n\rightarrow \infty } \int \! dV \: P_n(V)\, V^m \; .
\end{equation}
Once again, taking the $m$-th moment and taking the limit in $n$ do not commute because of the absence of strong convergence. The addition of the $\delta$-function in eq.~(\ref{deltafunction}) is a  formal way of achieving strong convergence.
The situation is somewhat subtler for the infinite chain, where we know that different moments
diverge at different values of $p$. This cannot be reproduced by adding a single $\delta$-function
contribution at infinity. Instead, it may be that when the $m$-th moment $\langle V^m \rangle$
diverges, the probability distribution acquires a term $\propto \delta^{(m)}(V-\infty)$.

\section{Generalization to realistic models}

So far we assumed that the parameters $H$ and $\dot\phi$ do not depend
on the position along the potential. Although in the slow-roll
approximation these parameters are slowly varying, {\em i.e.}~they are
approximately constant during an Hubble time, one cannot neglect their
variation over a large range of $\phi$~\footnote{Some specific realistic models with potential of the form $\phi^n$ were first studied in the context of eternal inflation in \cite{Linde:1986fd,Goncharov:1987ir}, reaching conclusions qualitatively similar to ours.}. Let us discuss how our results
change taking this effect into account. We want to prove the
following. {\em We have eternal inflation, i.e.~a non-zero probability
  of an infinite reheating volume, if and only if $\Omega < 1$
  somewhere along the potential. } This statement is true up to
slow-roll corrections  and up to corrections of order
$(H/\Delta\phi)^2$, where $\Delta\phi$ is the range of $\phi$ where
the slow-roll conditions are satisfied.

The proof is quite easy. Consider the case in which $\Omega <1 $
somewhere along the potential. The slow-roll conditions imply that one
can take a range of $\phi$ around this point, with $\Delta\phi \gg
\dot\phi \cdot H^{-1} \sim \Omega^{1/2} H$, in which $\Omega$ and $H$
can be treated as constant. Now the idea is to restrict to this
interval and prove that the extinction probability is less than
one. For example one can consider a model with an absorbing barrier at
the downhill limit of the interval and with a reflecting one at the
uphill limit. This model is studied explicitly in Appendix
\ref{app:moments}, where it is shown that the extinction probability
gets less than one when $\Omega <1$, up to corrections ${\cal
  O}(H/\Delta\phi)^2$. On the other hand, this restricted model has
clearly an extinction probability larger than the model probing the
full potential. This proves that $\Omega <1 $ ensures an extinction
probability strictly less than 1, {\em i.e.}~a non-zero probability of
an infinite reheating volume. 

The other case is even simpler. If $\Omega > \Omega_0 > 1$ everywhere,
then one can consider a model with $\Omega = \Omega_0$
everywhere. This model has an extinction probability smaller than the
original one. On the other hand we know that the extinction
probability with constant $\Omega >1$  is 1. This proves that the extinction probability is 1 if
$\Omega >1$ everywhere: the
probability of infinite reheating volume vanishes.
We see that the condition for eternal inflation is local in field
space: it is enough to have $\Omega <1$ in a range of field values
parametrically larger than $H$. 

This also shows that all our conclusions do not change if we assume
that the last stage of inflation before reheating is far from the
eternal regime, {\em i.e.} $\zeta \ll 1$, as we observe for example in our
Universe. The additional e-folds of inflation with negligible
perturbations act as a sort of physical smoothing of the reheating
surface: $\zeta$ is large only on scales which are huge compared to
the horizon at the end of inflation. This is the way one should
understand the somewhat artificial smoothing of the reheating surface
we used in the paper.

% Everything we said so far will receive corrections proportional to the
% slow-roll parameters. One possibility is that the transition to
% eternal inflation remains sharp even taking into account these
% corrections; only the threshold being slightly modified. However to
% study these 
% corrections one has to consider also self-interactions among the
% inflaton perturbations as discussed in section
% \ref{sec:perturbative}. The simple picture of random walks and its
% analogy with branching processes changes and
% this may lead to qualitative differences. For example it is not
% obvious that all momenta will start diverging at the same point where
% also the extinction probability becomes less than 1.

It is interesting also to see whether our results can be extended to non-minimal models of single field inflation.  This will include k-inflation models \cite{Garriga:1999vw} with reduced speed of sound, like DBI inflation \cite{Alishahiha:2004eh}, and models where higher derivative terms are important, like ghost-inflation \cite{ArkaniHamed:2003uz,Senatore:2004rj}. A useful way of writing down the most generic theory of single field inflation has been recently studied in \cite{Creminelli:2006xe, Cheung:2007st}, where modifications with respect to the minimal slow-roll scenario are parametrized in terms of operators for the perturbations around the inflating background. 

There is an important qualitative difference between these non-minimal
models and the simplest slow-roll inflation case: {\em eternal
  inflation may lie outside the regime of validity of the effective
  field theory}. In the case of slow-roll, we discussed in section
\ref{sec:perturbative} that, even in the eternal inflation regime
$\zeta \sim 1$,  non-linearities are small and can be treated as
corrections to the free field picture \footnote{Even in the slow-roll
  case non-linearities become important if one is interested in very
  large (and very unlikely) fluctuations $\zeta \gg 1$. In particular
  the quartic interaction eq.~(\ref{eq:S4}) will become as important as the
  quadratic action for $\delta\phi \gtrsim H/\sqrt{\epsilon}$. It is
  easy to realize however the these extremely unlikely fluctuations
  are irrelevant in all our discussion. What must be weakly coupled
  are the typical fluctuations, which means $\zeta \sim 1$ when one
  gets close to the eternal inflation regime.}. However this is not the case for more general models. Eternal inflation can be studied only if the theory is weakly coupled for $\zeta \sim 1$; otherwise this regime is sensitive to the UV completion and cannot be studied within the effective field theory. 

Whether the theory is weakly coupled or not at $\zeta \sim 1$ will depend on the size of the various operators, but some general conclusion can be drawn. For models with a reduced speed of sound, $c_s < 1$, it is shown in \cite{Cheung:2007st}, that the same operator that reduces the speed of perturbations also induces interactions among fluctuations. The relation between the two features is fixed by the symmetries of the system. In particular one can show that the effect of cubic and quartic interactions at horizon crossing with respect to the quadratic action is of order \cite{Seery:2005wm,Chen:2006nt,Huang:2006eh,Cheung:2007st}
\be
\frac{S_3}{S_2} \sim \left(\frac{1}{c_s^2} -1 \right)\zeta \qquad \frac{S_4}{S_2} \sim \left(\frac{1}{c_s^2}-1\right)^2 \zeta^2 \;,
\ee
which should be compared with eq.s (\ref{eq:S3}) and (\ref{eq:S4}) in
the case of slow-roll. This implies that the theory becomes strongly
coupled at horizon crossing if one approaches the eternal inflation
regime $\zeta \sim 1$, unless $c_s$ is very close to one
\cite{ArkaniHamed:2007ky}. In other words eternal inflation is out of
the regime of validity of the effective field theory for models where
the speed of perturbations deviates substantially from one.  Similar conclusions were reached in the context of brane inflation in \cite{Chen:2006hs} and in the context of models with small speed of sound in  \cite{Leblond:2008gg}.

The same conclusion holds in another interesting case of single field
inflation: that of ghost inflation \cite{ArkaniHamed:2003uz,Senatore:2004rj}.  Indeed the Lagrangian for the
canonically normalized perturbation $\pi$ contains an interaction of the form
$\dot\pi^3/2 M^2$ where $M^2=\dot\phi$ is the velocity of the time
dependent condensate. Again comparing this with the quadratic
Lagrangian
\be
\frac{S_3}{S_2} \simeq \frac{\dot\pi^3/M^2}{\dot\pi^2} \simeq \frac{H
  \pi}{\dot\phi} \simeq \zeta \;.
\ee
This implies that the theory is strongly coupled in the regime of
eternal inflation $\zeta \sim 1$.

Exactly the same argument holds for the other de Sitter limit of
inflation discussed in \cite{Cheung:2007st} where, using the language
of ref.~\cite{Cheung:2007st}, the unitary gauge
operator $(g^{00}+1) \delta K^\mu {}_\mu$ dominates at horizon
crossing. Also in this case the theory contains the cubic operator
$\dot\pi^3/2 M^2$, so that again eternal inflation is not under control
of the effective field theory.

\section{Summary and discussion}
In this paper we gave a precise definition of slow-roll eternal
inflation by identifying a sharp change of behaviour of the system
at the critical value 
\be
\frac{\dot\phi^2}{H^4} = \frac{3}{\,2 \pi^2} \;.
\ee
A model gives rise to eternal inflation if and only if $\Omega \equiv
2 \pi^2 \dot\phi^2/(3 H^4) <1$.

We reviewed in Sec.~\ref{sec:intro} the reasons why inflation is under
control even in the eternal regime. It is easy to check that the system can be
perturbatively studied in an expansion
in the slow-roll parameters and in $H^2/M_{\rm Pl}^2$; indeed these small
quantities suppress both the deviation of the metric from exact de
Sitter and the self-interactions of the scalar degree of freedom. Our
analysis has been done at leading order in these parameters.

Starting from a finite inflationary volume, we identified the volume of the
Universe at reheating, smoothed on scales much larger than the Hubble
radius $H^{-1}$, as indicator of the onset of eternal
inflation. We expect, in fact, the reheating volume to become larger
and larger without bound as we approach the eternal regime.  Indeed in
Sec.~\ref{sec:volume} we
proved, at leading order in slow-roll and $H^2/M_{\rm Pl}^2$, that all moments
of the reheating volume probability distribution $\rho(V)$ diverge at
$\Omega =1$, signaling a sharp change in the behaviour of the system. 

To get further insight into the nature of this transition, we studied in
Sec.~\ref{sec:bacteria} a
discretized version of slow-roll inflation. The dynamics of inflation
can be mimicked, with a biological analogy, with a set of bacteria
which can hop left or right on a one dimensional lattice, representing
the position of the inflaton. The expansion is represented by the
constant reproduction rate of these bacteria, which at every time step
split into a fixed number of independent offsprings. This discrete version of the
system is simpler to study, as one can apply to it the whole machinery
of the theory of branching processes. On the other hand it can be
shown to exactly approach the continuous case in the limit in which
the lattice spacing goes to zero. This discrete model allows us to
reach the conclusion that the onset of eternal inflation at $\Omega
=1$ corresponds to the development of a non-zero probability of strictly infinite
reheating volume: this is the sharpest definition of eternal inflation.

Our results are obtained at leading order in slow-roll and
$H^2/M_{\rm Pl}^2$. Even small corrections to the asymptotic time
dependence would completely change the results; for example this would
happen if the reheating probability (\ref{eq:prexp}) goes at large
times as $\exp(t^{1+\epsilon})$ where $\epsilon$ is small
correction \footnote{We thank David Gross for emphasizing this point.}. 
From a preliminary analysis, we expect that this is not the case and that the qualitative picture of a sharp
transition remains unaltered, with only perturbative corrections to
the exact point where the transition happens. The proof of this,
however, is not entirely straightforward. Indeed, at subleading order in our expansion
parameters many new ingredients must be taken into account. The Hubble
rate $H$ and inflaton speed $\dot\phi$ are not constant anymore, but depend on the
position along the potential; the scalar degree of freedom is not a
free field anymore as cubic and quartic self-interactions must be
taken into account; the metric deviates from exact de Sitter and,
finally, also tensor modes must be taken into account. We expect the
diffusion equation to still capture the relevant dynamics of the
phase transition, but further work is needed to extend our
results at subleading order. Another weak point of our discussion is
that eternal inflation is characterized studying a classical, smoothed
observable: 
the reheating volume. Although the exact value of the smoothing scale
does not appear in the final equations (in the limit in which it is
much larger than the Hubble radius $H^{-1}$), it would be nice to have
a more intrisic and quantum mechanical definition of eternal
inflation, without need of a smoothing procedure.

The sharp characterization we gave of slow-roll eternal inflation may
be a first step towards a more profound understanding of its features
and implications. In particular there are reasons to suspect that the
semiclassical description of eternal inflation, with its infinite
creation of volume may be misleading (see for example
\cite{ArkaniHamed:2007ky}) or even that eternal
inflation itself may be censored in a fundamental theory \cite{ArkaniHamed:2008ym}; if this is the case
the effective theory description must break down before $\Omega = 1$.  

In the context of the conjectured dS/CFT correspondence put forward in \cite{Strominger:2001pn}, it would be
interesting to understand what is the nature of the transition at
$\Omega =1$ on the CFT side. Another possible direction is to try to
understand the connection between our critical point $\Omega =1$ and
other critical point for the existence of eternal inflation, like in
the case of old inflation \cite{Guth:1982pn} and topological inflation \cite{Vilenkin:1994pv,Linde:1994hy}.

\section*{Acknowledgments}
It is a pleasure to thank Nima Arkani-Hamed, Lotfi Boubekeur, Raphael
Bousso, Ben Freivogel, Andrea Gambassi, Jaume Garriga, David Gross,
Alan Guth, Andrei Linde, Matt Kleban,
Raman Sundrum and Filippo Vernizzi for useful discussions and comments.

\appendix

\section*{Appendices}

\section{Proof that all moments diverge at the same $\Omega$}

\label{app:moments}
In this Appendix we want to prove that in the presence of a regulating
uphill barrier all the moments of the distribution $\rho(V)$ start
diverging at the same point $\Omega=1$.

Let us first solve the diffusion equation with a drift term
\be
\frac{\partial P}{\partial t} = -\dot\phi \frac{\partial P}{\partial
  \phi} + \frac{H^3}{8 \pi^2} \frac{\partial^2 P}{\partial\phi^2}
\ee
in the presence of an absorbing barrier at $\phi = \phi_r$ and a
reflecting one at $\phi=\phi_{up}$. At the reheating barrier the
boundary condition is $P=0$, while on the uphill barrier we have
to impose a vanishing value of the probability current
\be
\dot\phi P - \frac{H^3}{8 \pi^2} \frac{\partial
  P}{\partial\phi} =0  \quad {\rm at} \quad \phi = \phi_{up} \;.
\ee
Using the standard method of separation of variables one can write a
general solution as
\be
\label{eq:general}
\exp\left[\frac{4\pi^2 \dot\phi\,\phi -2\pi^2 \dot\phi^2 t}{H^3}\right]
\sum_n a_n \sin \left[\frac{\alpha_n \pi (\phi_r - \phi)}{\phi_r-\phi_{up}}\right]
\exp\left[{-\frac{H^3 t}{8 \pi^2} \cdot \frac{\alpha_n^2
      \pi^2}{(\phi_r-\phi_{up})^2}}\right] \;.
\ee
The boundary condition at the reheating point is clearly satisfied, while the coefficients $\alpha_n$ must be chosen so that each sinus
satisfies the boundary condition on the uphill barrier. This gives the
following equation
\be
\label{eq:tglinear}
\tg (\alpha_n \pi) = - \frac{H^3}{4 \pi^2 \dot\phi} \cdot
\frac1{\phi_r-\phi_{up}} \cdot \alpha_n \pi \;.
\ee
Let us check first of all that, sending $\phi_{up} \to - \infty$, the result reproduces the case with only the reheating barrier. In the limit in which the uphill barrier is removed we see that the solutions of (\ref{eq:tglinear}) behaves as $\alpha_n \to
n$, so that the series reduces to a normal Fourier
decomposition. Using the Fourier decomposition of the $\delta$ function we take the coefficients of the series to be
\be
a_n = \frac{2}{\phi_r-\phi_{up}} \sin \frac{n \pi
  \phi_r}{\phi_r-\phi_{up}} \;.
\ee
Therefore the series in eq.~(\ref{eq:general}) takes the form
\be
\sum_n \exp\left[{-\frac{H^3 t}{8 \pi^2} \cdot \frac{n^2
      \pi^2}{(\phi_r-\phi_{up})^2}}\right] \frac{1}{\phi_r-\phi_{up}} 
      \left[\cos\frac{n \pi \phi}{\phi_r-\phi_{up}} - \cos\frac{n \pi
          (2 \phi_r- \phi)}{\phi_r-\phi_{up}} \right] \;.
\ee
In the limit $\phi_{up} \to - \infty$ the sum becomes an integral. The
first sum of cosines becomes 
\be
\int_0^{+ \infty} \!\!\!\!dX \,e^{-\frac{H^3 t}{8} X^2} \cos{(\pi X \phi)} \sqrt{\frac{2 \pi}{H^3 t}} e^{-\frac{2 \pi^2 \phi^2}{H^3 t}} \;.
\ee
Taking into account the exponential factor in front of the series in eq.~(\ref{eq:general}), we see that we reproduce the result (\ref{image}) in the presence of the reheating barrier only. 

The important feature for us of the general solution (\ref{eq:general}) is that at late times it is dominated by the first term of the sum, which is the slowest to decay. For large and negative $\phi_{up}$, $\alpha_1 \simeq 1$ so that the full time dependence of the leading term is
\be
\label{eq:simple}
\exp \left[-\frac{2\pi^2 \dot\phi^2 t}{H^3}-\frac{H^3 t}{8
    (\phi_r-\phi_{up})^2}\right] \sim \exp \left[-\frac{2\pi^2
    \dot\phi^2 t}{H^3}\right] = \exp (- 3 H \Omega t)\;.
\ee

We are interested in studying the $n$-th moment of the reheating volume
distribution, which is easily related to the $n$-point joint reheating probability $p_{nr}$ by
\be
\label{eq:moments}
\langle V^n \rangle =  \int dx_1 \ldots dx_n \int dt_1 \ldots dt_n\,
e^{3H (t_1 + \ldots + t_n)}\,  p_{nr}(t_1, \ldots, t_n, \vec x_1,
\ldots, \vec x_n) \;. 
\ee
Clearly, if $\langle V \rangle$ diverges, also  $\langle V^n \rangle$
diverges, therefore we concentrate on the case $\Omega >1$, when $\langle V \rangle$ is finite.

The simple time dependence of eq.~(\ref{eq:simple}) will also appear in the reheating
probability. Let us take $n$ points. Going backwards in time they progressively merge together: at $t_{* (n-1)}$ the distance between the two closest points gets smaller than $H^{-1}$ and we are left with $n-1$ independent points; at $t_{*(n-2)}$ a couple of the remaining points merge to $n-2$ and the process continues until $t_{*1}$, when we are left with a single point.  It is easy to prove by induction from 
eq.~(\ref{eq:simple}) that  
\be
\label{eq:pnr}
p_{nr}(t_1,\ldots t_n, \vec x_1, \ldots, \vec x_n) \sim \exp \left[-3H\Omega (t_1+t_2 + \ldots + t_n) + 3H\Omega (t_{*1}+ t_{*2}+ \ldots + t_{*(n-1)}) \right]  \;.
\ee

Let us now use this expression to study the convergence of $\langle V^n \rangle$, eq.~(\ref{eq:moments}). For a given comoving position of the $n$ points, the integral
over time can be divided into many intervals each going from $t_{min}$ to $t_{max}$; in each interval the $n$ points are grouped into 
$m$ independent subsets. At very early times all the points are
separated by a physical distance $\ll H^{-1}$, so that their evolution
is completely correlated, so that they form a single group. As time passes the $n$ points split in
subgroups: the elements of each subgroup still evolve together, while
the evolution of different subgroups is not correlated \footnote{As in the case
of the variance we neglect the intervals of partial correlation using
a two-step approximation.}. We obtain
\be
\int dx_1 \ldots dx_n \sum_{intervals} \int^{t_{max}}_{t_{min}} dt_1
\ldots dt_m e^{3H (t_1 \cdot k_1 + \ldots + t_m \cdot k_m)}
p_{mr}(t_1, \ldots, t_m, \vec x_1,
\ldots, \vec x_m) \;,
\ee
where the sum is extended over the different time intervals during which the $n$
points are divided into $m$ groups, each group consisting of $k_i$ points.  Before $t_{min}$
some points belonging to different groups are correlated (using the notation of the previous paragraph $t_{min}=t_{*(m-1)}$), after
$t_{max}$ some of the points in the same group start to evolve
independently as they become separated by more than $\sim H^{-1}$. Clearly
the extrema $t_{min}$ and $t_{max}$ are functions of
the comoving coordinates $x_1 \ldots x_n$.          

Using eq.~(\ref{eq:pnr}) we can write 
\be
\langle V^n \rangle =\int dx_1 \ldots dx_n \sum_{intervals} e^{3 H \Omega (t_{*1} + \ldots + t_{*m})} \int^{t_{max}}_{t_{min}} dt_1
\ldots dt_m e^{3H(k_1-\Omega) t_1} \ldots e^{3H(k_m-\Omega) t_m} \;.
\ee
The next step is to rewrite the integration over comoving coordinates as integrals over time variables, similarly to what we did in
eq.~(\ref{saddleV2}) for the $\langle V^2 \rangle$ case. 
The distances among the various groups fix the values of $t_{*i}$. For a particular sequence of merging of the groups going backwards in time,  one can choose the following spatial coordinates:
$e^{-Ht_{*(m-1)}}$ is the distance between the two closest groups, $e^{-Ht_{*(m-2)}}$ is the distance between the next to closest and so on. 

One can thus perform all the integrals over the relative positions of the groups, except for the one over $t_{min}$: 
\be
\prod_i \int_{t_{*(i-1)}}^ {t_{*(i+1)}} \!\!\! dt_{*i} \, e^{3H(\Omega-1) t_{*i}} \sim  e^{3H(\Omega-1)(m-1)t_{min}} \;,
\ee 
where we used that for $\Omega >1$ each integral is dominated by the upper limit of integration.
The integral over the positions of the points inside each group can be done noticing that $t_{max}$ fixes the maximum distance inside any group: the points with the largest separation will be the first ones to start an independent evolution, changing the way the $n$ points are divided. Thus the integral over the internal coordinates give   
\be
\sim \prod_{i=1}^m  e^{-3 H (k_i-1) t_{max}} \;.
\ee
Including $t_{min}$ and $t_{max}$, this exhausts all the spatial coordinates except one, whose integral just gives an overall comoving volume $L^3$. 

For each time interval we are left with 
\be
\int dt_{min} \int dt_{max} \, e^{3 H (\Omega-1) (m-1) t_{min}}
\prod_{i=1}^m e^{-3 H (k_i-1) t_{max}}  \prod_{j=1}^m\int_{t_{min}}^{t_{max}} \!\!\!\! dt_j \,
e^{3H(k_j-\Omega)t_j} \;.
\ee
Each of the integrals in the product at the end gives
\begin{eqnarray}
\sim e^{3H(k_j-\Omega) t_{max}} \quad {\rm for}\; k_i > \Omega && \\
\sim e^{3H(k_j-\Omega) t_{min}} \quad {\rm for}\; k_i < \Omega 
\end{eqnarray}
For $\Omega >1$ the integral over $t_{max}$ is always convergent and
dominated by its lower bound $t_{min}$. Therefore we end up with a
simple integral over $t_{min}$
\be
\int dt_{min} \, e^{-3H (\Omega-1) t_{min}} \;,
\ee
which is always convergent for $\Omega >1$.

This proves that $\langle V^n \rangle$ always converges when $\langle
V\rangle$ does and therefore that all moments start diverging at the same
critical point $\Omega =1$.

\section{Asymptotics of  $\langle 1|(S+S^\dag)^{m}|i\rangle$}
\label{sheets}
Here we will prove the estimate (\ref{binom_estimate}) for the leading exponential asymptotics for
matrix elements of the form  $A^{(m,i)}\equiv\langle 1|(S+S^\dag)^{m}|i\rangle$ at large $i$.
Let us start with deriving a recursion relation for these matrix elements in analogy to (\ref{recursive1}).
One writes
\begin{eqnarray}
\nonumber
A^{(m,i)} & = &\langle1|\left(S+ S^\dag \right)^{m-1}\left(S+S^\dag\right)|i\rangle \\
& = &  A^{(m-1,i-1)}+A^{(m-2,i)}A^{(0,1)} + \langle1|\left(S+
S^\dag \right)^{m-2}\left(S+S^\dag\right)S |1\rangle   \nonumber \\
 & = & \! \ldots \; = A^{(m-1,i-1)}+\sum_{k=0}^{m-2}A^{(k,i)}A^{(m-2-k,1)} \; .
 \label{recursionmi}
\end{eqnarray}
As before, it is convenient to introduce a generating function
\[
G(x,y)=\sum_{m=0}^\infty\sum_{i=1}^\infty A^{(m,i)}x^my^i\;
\]
It is straightforward to check that the recursion relation (\ref{recursionmi}) and $A^{(0,1)}=1$ 
implies the following
algebraic equation for $G$,
\[
xy \, G(x,y)+x^2G(x,y)F(x^2)=G(x,y)-y\;,
\]
where $F(x)$ is the generating function (\ref{Fx}) for $A^{(m,1)}$. Consequently,
\be
\label{Gxy}
G(x,y)={2y\over 1+\sqrt{1-4x^2}-2xy}\;.
\ee
Unlike for the function $F$, finding the Taylor coefficients of $G$ in closed 
form is rather challenging. However, what we need is just the leading exponential asymptotics 
of $A^{(\mu  i,i)}$ at large $i$ and fixed $\mu$. In other words, we are looking for $\lambda$, such that
\be
\label{lambdadef}
A^{(\mu   i,i)}=\lambda^if(i)
\ee
where $f(i)$ is not exponential at large $i$. To find this asymptotics it is convenient
to work with another pair of variables $(x,z=y\, x^\mu)$, so that
\be
\label{Gxz}
G(x,z)=\sum_{m=0}^\infty\sum_{i=1}^\infty A^{(m,i)}x^{m-\mu i}z^i\;\;.
\ee
Then the coefficients $A^{(\mu i,i)}$ are generated by $G_0(z)$, which is 
the $x$-independent part of $G(x,z)$,
\be
\label{G0}
G_0(z)=\sum_{i=1}^\infty A^{(\mu i,i)}z^i
\ee
Note that $G_0(z) \neq G(x=0,z)$, because
the series for $G(x,z)$ contains negative powers of $x$. Instead, it is given by the contour integral
\be
\label{contour}
G_0(z)=\oint{dx\over x}G(x,z)=\oint{dx\over x^2}{2z\over x^{\mu-1}(1+\sqrt{1-4x^2})-2z}
\ee
in the complex $x$-plane.
Here the integration contour should enclose the point $x=0$, and should not exit the region
where the series (\ref{Gxz}) converges, so that one can interchange the order of summation and integration. From the definition (\ref{Gxy}) it follows that this is the case for $x\ll1$ and $y=z/x^\mu\ll 1$.

To find the value of $\lambda$ in (\ref{lambdadef}) note that  $G_0(z)$, by its definition (\ref{G0}), is analytic for
$| z |< | \lambda|^{-1}$ and diverges at $| z | = |\lambda|^{-1}$. Given that all coefficients in (\ref{G0}) are real positive numbers, we can
restrict to real values of z. The integrand in
(\ref{contour}) has two cuts extending from $x=\pm 1/2$ to $x=\pm \infty$, and a double pole at $x=0$. Also at small $z$ it has $(\mu-1)$ additional poles near $x=0$.
The integration contour encloses all these poles and does not cross the cuts.
At larger values of  $z$ one has two extra poles entering from the second sheet---they appear in the first sheet at $x=\pm 1/2$ when $z$ crosses $1/2^\mu$. These poles are outside the integration contour. The singularity in $G_0(z)$ arises when, for even larger $z$, one of these two new poles merges with one of $(\mu+1)$ poles
inside the contour: at this value of $z$ the contour necessary passes through a singularity.
Consequently, a singularity appears when the denominator in (\ref{contour}) develops a double zero
(other than the trivial one at $x=0$). By calculating its derivative we find that this happens at 
\[
z=z_c\equiv 2^{-\mu}\mu^{-\mu}\l{\mu^2-1}\r^{\mu/2}\sqrt{\mu+1\over\mu-1}\;.
\]
 Hence $\lambda=z_c^{-1}$, exactly reproducing the estimate (\ref{binom_estimate}).
%%%%%%%%%%%%%%%%%%%%%%%%%%%%%%%%%
%%%%%%%%%%%%%%%%%%%%%%%%%%%%%%%%%

\end{document}